\documentclass[nolinenumbers]{aastex631}

\shorttitle{Physical parameters of contact binaries}
\shortauthors{Li et al.}

\graphicspath{{./}{figures/}}

\begin{document}

\title{Physical Parameters of 146 Contact Binaries Derived from Light and Radial Velocity Curves}


\author{Kai Li}
\correspondingauthor{Kai Li}
\email{kaili@sdu.edu.cn}
\author{Xiang Gao}
\author{Si-Rui Wang}
\author{Li-Heng Wang}
\affiliation{Shandong Key Laboratory of Space Environment and Exploration Technology, Institute of Space Sciences, School of Space Science and Technology, Shandong University, Shandong, China}

\begin{abstract}
We present a comprehensive analysis of 146 contact binaries using medium-resolution LAMOST spectra and photometric data from the All-Sky Automated Survey for SuperNovae (ASAS-SN) and the Transiting Exoplanet Survey Satellite (TESS). Radial velocity curves obtained through the cross-correlation function method were modeled simultaneously with the light curves using the Wilson–Devinney code to derive the physical parameters of these systems. The reliability of our results was verified through comparison with previous studies of ten systems, showing good agreement. Our analysis shows that the more massive components are generally less-evolved main-sequence stars, whereas the less massive components tend to be over-sized and over-luminous, consistent with earlier findings. The distribution of orbital angular momentum supports the scenario in which contact binaries form from detached binaries via angular momentum loss. We identified 38 low mass ratio ($q < 0.25$) systems, 11 of which have extremely low mass ratios ($q < 0.15$). A remarkable example is ASASSN-V J111451.48+005038.6, which exhibits a mass ratio of 0.113 and the highest fill-out factor (98.3\%) reported to date, making it a strong candidate for future mergers.  Conversely, we also identified 11 high mass ratio (H-type) systems, including ASASSN-V J093921.74+390452.6 - the system with the highest spectroscopically confirmed mass ratio ($q = 0.993$) and a low fill-out factor (1.8\%), suggesting it recently entered the contact phase.
Additionally, several empirical relations between physical parameters are established.
\end{abstract}


\section{Introduction} \label{sec:intro}
Contact binaries, also known as W UMa type variables, usually contain two late-type Roche lobe overfilling component stars. These systems are highly prevalent, having been identified not only in the field but also in both open and globular clusters \citep{1993ASPC...53..164K}. The light variation of contact binaries is typical EW-type, continuous variation and two nearly equal depth minima, indicating that the temperatures of the two component stars are almost identical despite very different component masses (mass ratios are usually less than 0.5, \citealt{2021ApJS..254...10L}). This may be caused by mass and energy transfer through the common envelop \citep{1968ApJ...151.1123L}. Contact binaries generally have orbital periods shorter than one day, with a pronounced short-period cutoff around 0.22 days \citep{2017RAA....17...87Q,2019MNRAS.485.4588L}. They also follow well-defined period–color \citep{1967MmRAS..70..111E} and period–luminosity–color relations (e.g., \citealt{1994PASP..106..462R,2016ApJ...832..138C,2018ApJ...859..140C}), making them excellent distance indicators.

Based on more than five decades of research, it has been proposed that contact binaries form from close detached binaries through angular momentum loss (AML) driven by magnetic braking (e.g., \citealt{1968ApJ...151.1123L,1988ASIC..241..345G,1994ASPC...56..228B,2005ApJ...629.1055Y,2006AcA....56..347S,2017RAA....17...87Q}), and they are expected to eventually merge into rapidly rotating single stars, and result in luminous red nova event \citep{2011A&A...528A.114T,2016RAA....16...68Z}. Nevertheless, the detailed formation mechanism of contact binaries remains an open question. Beyond that, there are many interesting phenomena of contact binaries that no theoretical model can superiorly and perfectly explain, such as the O'Connell effect (e.g., \citealt{1951PRCO....2...85O,2003ChJAA...3..142L}), the thermal relaxation oscillation theory (e.g., \citealt{1976ApJ...205..217F,2003MNRAS.342.1260Q}), the low mass ratio limit (e.g., \citealt{1995ApJ...444L..41R,2006MNRAS.369.2001L,2021MNRAS.501..229W,2022AJ....164..202L,2024A&A...692L...4L}), the very sharp short-period cut-off (e.g., \citealt{1992AJ....103..960R,2006AcA....56..347S,2019MNRAS.485.4588L}). \cite{1970VA.....12..217B} divided contact binaries into two types: A-type and W-type. For A-type systems, the more massive component has the higher temperature, while for W-type systems, it is on the contrary. Many studies tried to derive the connection and difference between these two type contact binaries, no clear conclusion has been determined at present (e.g., \citealt{2020MNRAS.492.4112Z}).
Progress in resolving these outstanding questions is currently hindered by the lack of precise and homogeneous absolute parameters (masses, radii, and luminosities) for a statistically significant sample of systems. Consequently, obtaining such precise fundamental parameters is a critical prerequisite for advancing theoretical models. 

Determining the absolute parameters of contact binaries requires both photometric light curves (LCs) and radial velocity (RV) curves. While large-scale photometric surveys (e.g., \citealt{2015JATIS...1a4003R, 2018MNRAS.477.3145J, 2020ApJS..249...18C, 2022ApJS..258...16P}) have enabled the discovery of thousands of contact binaries and provided easy access to high-precision LCs, the scarcity of RV data remains a major obstacle to deriving absolute parameters for the majority of these systems.
According to the International Variable Star Index (VSX\footnote{\url{http://www.aavso.org/vsx/}}; version 19 October 2025), more than 900,000 contact binaries have been identified to date. Systematic RV observations by Prof. Rucinski and collaborators (e.g., \citealt{1999AJ....118..515L, 1999AJ....118.2451R, 2000AJ....120.1133R, 2001AJ....122..402L, 2001AJ....122.1974R}) have yielded precise RV curves for 111 systems, significantly advancing the field. Nevertheless, this sample represents only a tiny fraction of the known contact binaries. The RV observations of contact binaries are still very insufficient. The Large Sky Area Multi-Object filber Spectroscopic Telescope (LAMOST) \citep{2012RAA....12.1197C,2015RAA....15.1095L,2020arXiv200507210L} started the medium-resolution spectroscopic (MRS) survey in September 2017. More than three million stars have been observed, which offers a unique opportunity to obtain the RV curves of a large sample of contact binaries.

This paper studies contact binaries that have been simultaneously observed by LAMOST MRS survey and All-Sky Automated Survey for SuperNovae (ASAS-SN; \citealt{2014ApJ...788...48S,2018MNRAS.477.3145J}).
We used the LAMOST MRS data to determine RV curves, which were then combined with the photometric LCs to derive their physical parameters.

\section{Observations} \label{sec:observations}
\subsection{Photometric Observations\label{sec:Photometric}}
ASAS-SN is a ground-based photometric survey operated by Las Cumbres Observatory \citep{2013PASP..125.1031B}.
Initiated in 2013, it monitors the entire sky down to V$\sim$17.0 mag with a cadence of 2–3 days, primarily to search for nearby bright supernovae and other transients. The survey comprises five stations located in Hawaii, Chile, Texas, and South Africa. Each ASAS-SN camera covers a field of view of approximately 4.5 deg$^2$ with a pixel scale of 8$^{\prime\prime}$.0. By arranging the cameras with moderately overlapping fields, the system achieves an instantaneous field of view of roughly 360 deg$^2$ across all stations. Observations were carried out using either V or g filters, and recorded three 90-second exposure images for every epoch. Under typical clear-sky conditions, ASAS-SN can survey more than 48000 deg$^2$ per night. Although its primary goal is the detection of nearby bright supernovae and other transient sources, the survey has also revealed a large number of variable sources. According to \cite{2018MNRAS.477.3145J} and \cite{2019MNRAS.486.1907J}, more than 420000 variable stars were identified in ASAS-SN V-band data between 2013 and 2018, including more than 70000 contact binaries.

The Transiting Exoplanet Survey Satellite (TESS; \citealt{2015JATIS...1a4003R,https://doi.org/10.17909/fwdt-2x66}) is an all-sky survey mission whose primary objective is the detection of Earth-sized exoplanets around stars brighter than 16 magnitudes. TESS observes the sky with a field of view of $24^\circ\times 96^\circ$, monitoring each sector for approximately 27.4 days. TESS provides high precision data with multiple cadences ranging from 2-minute to 30-minute. All TESS data are publicly accessible through Mikulski Archive for Space Telescopes (MAST)\footnote{\url{https://mast.stsci.edu/portal/Mashup/Clients/Mast/Portal.html}} .

\subsection{Spectroscopic Observations\label{sec:Spectroscopic}}
LAMOST is a 4-m reflecting Schmidt telescope equipped with 4000 fibers \citep{2012RAA....12.1197C,2015RAA....15.1095L,2020arXiv200507210L}. It features a field of view of approximately 5$^\circ$ and operates in two observational modes: low resolution and medium resolution. The low-resolution mode, initiated in October 2011, provides a spectral resolution of $R\sim1800$ and covers a wavelength range from $3700$ {\AA} to $9000$ {\AA}. In this mode, the targets span an r-band magnitude range of 9.0 to 17.5 mag. The medium-resolution mode, launched in September 2017, achieves a spectral resolution of $R\sim7500$, with wavelength coverages of $4950$ {\AA} to $5350$ {\AA} for the blue arm and from $6300$ {\AA} to $6800$ {\AA} for the red arm. For this mode, targets are selected within a G-band magnitude range of 9.0 to 15.0 mag. Studies such as those by \cite{2021ApJS..256...31L}, \cite{2022ApJS..258...26Z}, \cite{2024MNRAS.527..521K}, and \cite{2025ApJS..276...11L} have utilized LAMOST MRS data to identify double-lined spectroscopic binaries. In addition, \cite{2022AJ....163..235L} derived RV curves for two contact binaries and determined their absolute parameters. More recently, \cite{2025ApJ...979...69W} conducted a systemic search for and analysis of eclipsing binaries in the LAMOST MRS field. These efforts have motivated our initiative to extract RV curves for a large sample of contact binaries using LAMOST MRS data. In this study, we use the MRS data of LAMOST Data Release 9\footnote{\url{http://www.lamost.org/dr9/}} (DR9), which includes 8,226,434 spectra.

\section{Target selection and RV determination} \label{sec:TSandRV}
We used the MRS catalog of LAMOST DR9 to cross-match with the contact binary (EW type) catalog of ASAS-SN using a 3$^{\prime\prime}$ radius, resulting in 2519 systems.
To ensure a good phase coverage for the RV curves, we selected stars with at least five exposures.
The signal-to-noise ratio (SNR) was required to be greater than 10, as reliable RVs are difficult to obtain from low-SNR spectra. Applying these two criteria, 437 targets were selected for further analysis.

\begin{figure*}
\centering
\includegraphics[width=0.32\textwidth]{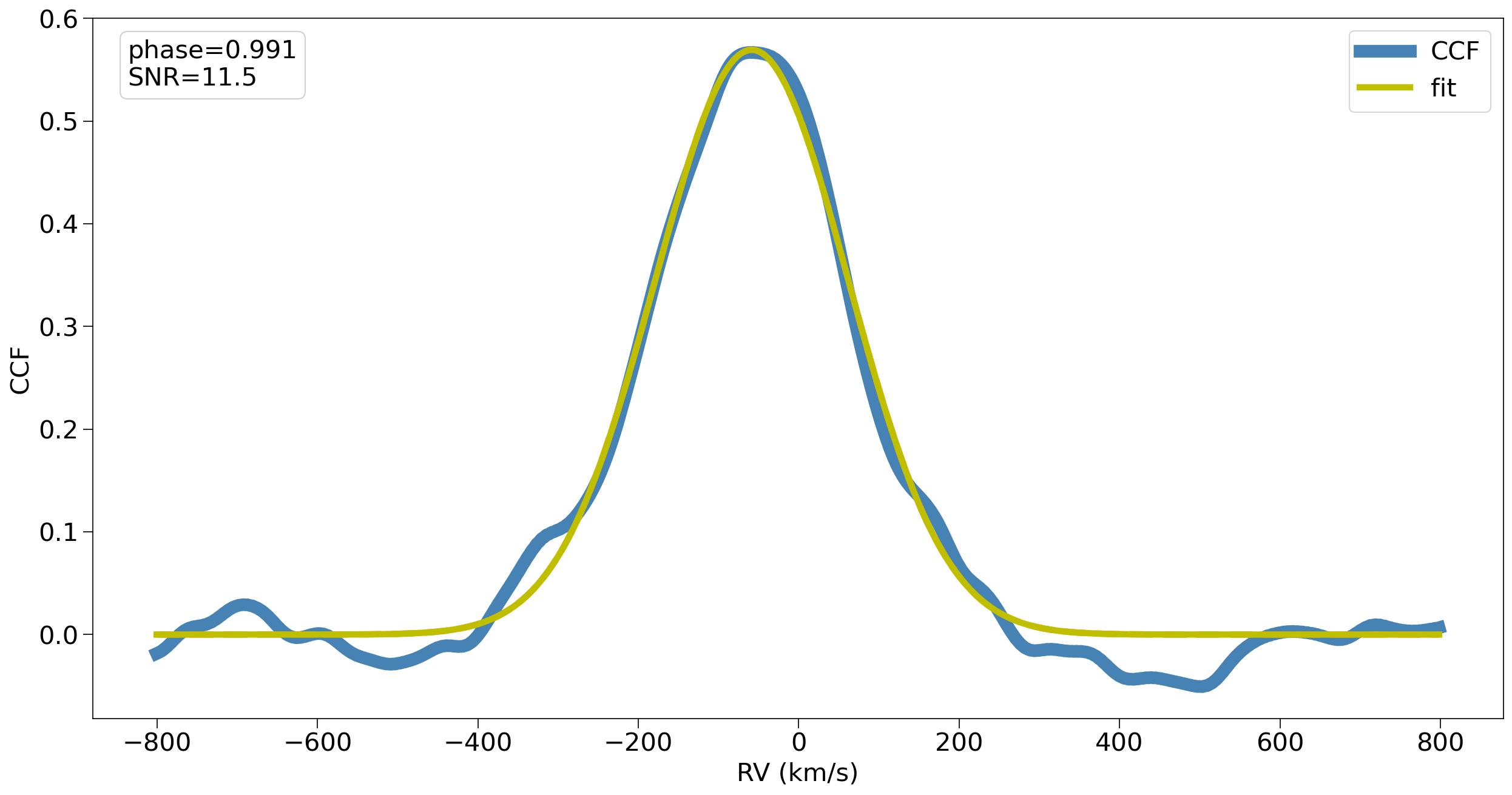}
\includegraphics[width=0.32\textwidth]{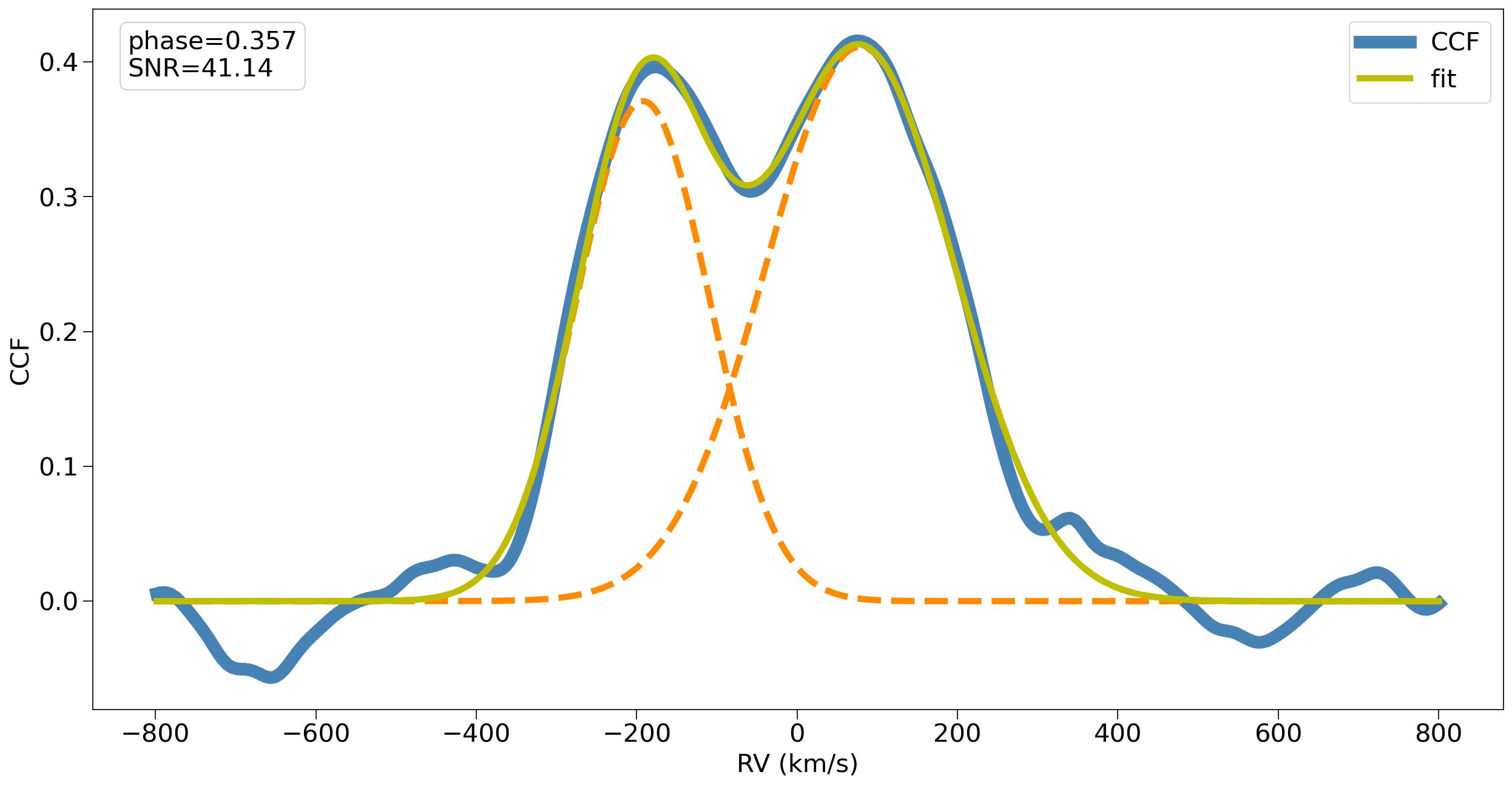}
\includegraphics[width=0.32\textwidth]{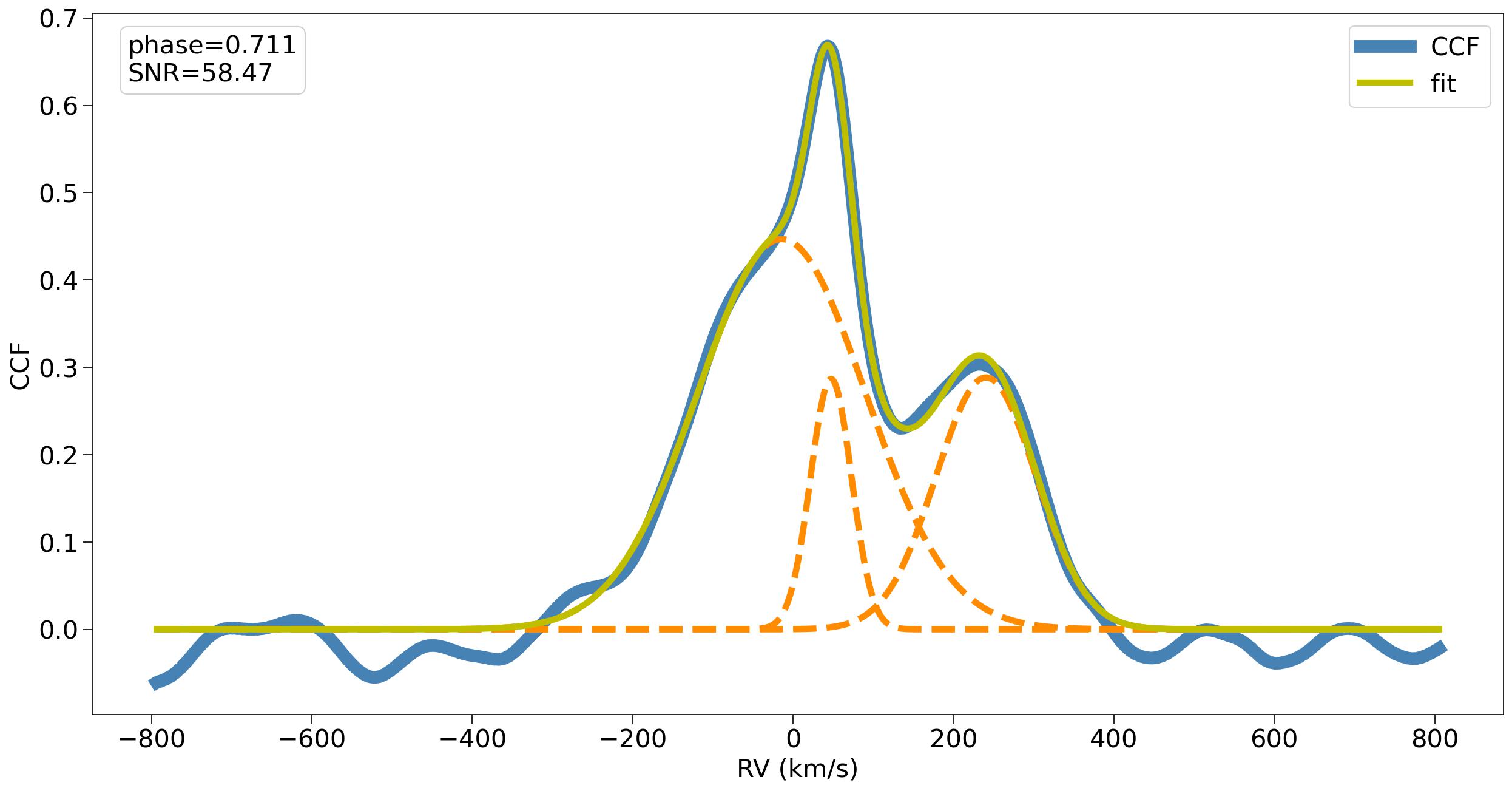}
\caption{The CCF curves and their fitting of three targets. The panels from left to right correspond to single, double, and triple Gaussian fits, respectively.}
\label{CCF}
\end{figure*}

Only the MRS data from the blue arm were adopted to measure RVs, as this region contains more absorption lines than the red arm, enabling RV measurements with a precision of a few km/s for most stars \citep{2021ApJS..256...31L}. The RVs of our targets were determined through cross-correlation function (CCF) analysis using PHOENIX synthetic spectra \citep{2013A&A...553A...6H} as templates. For the RV extraction, we employed PHOENIX model spectra with 10000 resolution, covering temperatures from 4000 to 10000 K, which optimally matches the temperatures of our target stars. RVs were determined by fitting the CCF profiles using GaussPy+ \citep{2019A&A...628A..78R}, a Python-based implementation of the Autonomous Gaussian Decomposition algorithm. The choice of the number of Gaussian components depends on the orbital phase. Near orbital phases 0 and 0.5 (conjunction), the radial velocities of the two components are nearly equal, resulting in a single-peaked CCF profile that can be adequately fitted with a single Gaussian. At other phases, the two components are separated in velocity space, producing a double-peaked CCF that requires a double-Gaussian fit. For a small number of targets showing a clear triple-peaked CCF, the third peak likely arises from a third light source (e.g., a tertiary companion or background star), necessitating a triple-Gaussian fit. Figure \ref{CCF} displays three representative examples of these CCF fits. All CCF fits and resulting RV curves were visually inspected, and targets with unreliable RV curves due to low SNR or poor phase coverage were discarded. Finally, 146 targets were retained, and their RVs are listed in Table \ref{tab:RV}. 

\begin{table}
\small
\centering
\caption{ The RVs of the 146 targets} 
\begin{tabular}{lllllll}
\hline
Star	 & time (BJD)	  &Phase	&RV$_1$ (km/s)	&Errors	&RV$_2$ (km/s)	&Errors\\ 
\hline
    
ASASSN-V J000916.24+360536.6	&2458831.94410 	&0.161	&-154.8	      &2.1	  &68.2	        &1.8   \\
&2458831.96031 	&0.208	&-179.2	      &1.8	  &68.5	        &2.0   \\                          
&2459130.11196 	&0.264	&-174.6	      &1.2	  &100.9	      &0.8   \\                         
&2458831.99293 	&0.303	&-160.2	      &2.0	  &88.2	        &1.3   \\                          
&2459130.12849 	&0.311	&-164.3	      &1.1	  &96.1	        &0.9   \\                         
&2459130.15555 	&0.390	&-129.1	      &2.2	  &76.4	        &2.0   \\                          
&2459130.17183 	&0.437	&-68.3	      &2.8	  &91.3	        &2.3   \\                          
&2459130.19083 	&0.492	&	-           &-      &-2.5	        &0.6   \\                          &2459130.20709 	&0.540	&	-           &-      & 0.1	        &0.9   \\                              &2458825.93725 	&0.733	&181.8	      &1.6	  &-106.8	      &1.4   \\  
\hline
\end{tabular}
\label{tab:RV}
Note. This table is available in its entirety in machine-readable form in the online article.
\end{table}

\section{Determination of physical parameters} \label{sec:physical parameters}
\begin{figure*}
\centering
\includegraphics[width=0.45\textwidth]{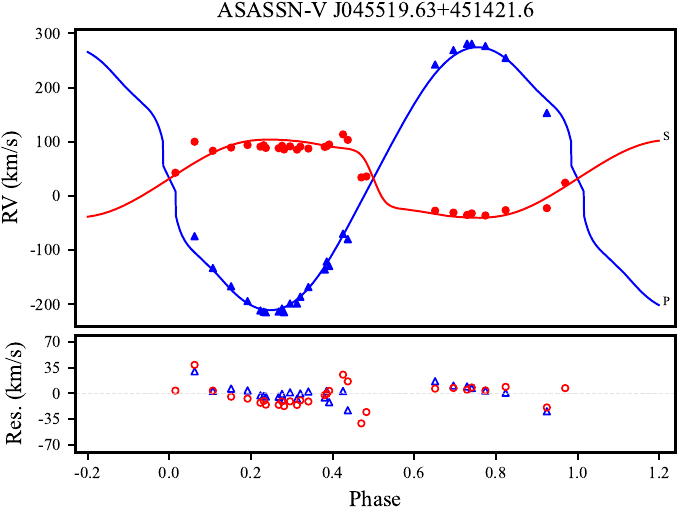}
\includegraphics[width=0.45\textwidth]{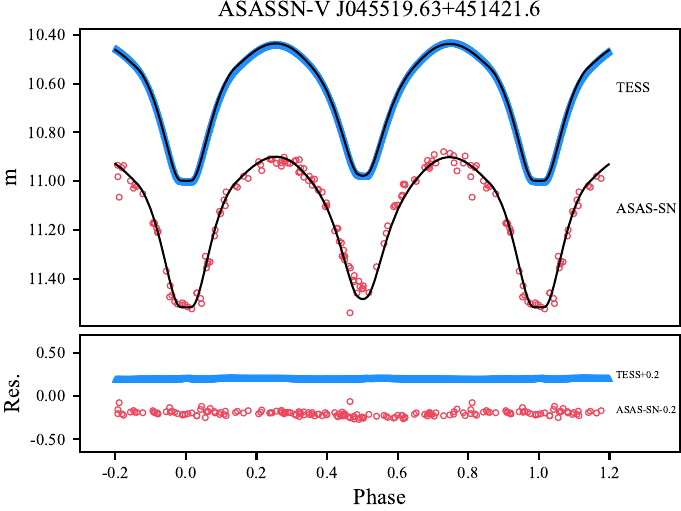}
\includegraphics[width=0.45\textwidth]{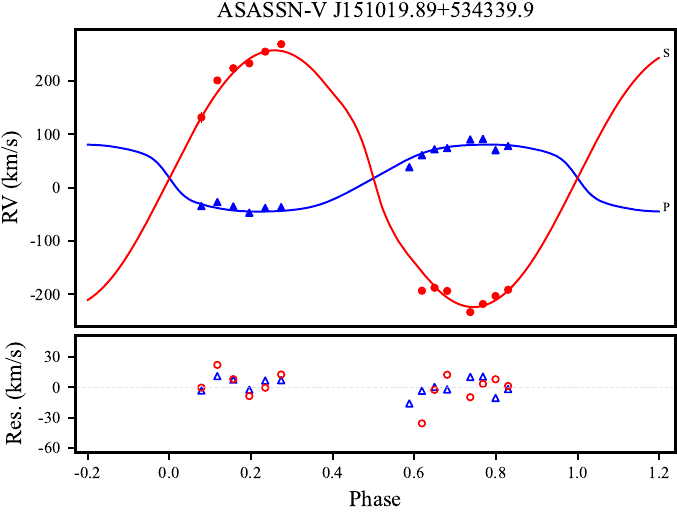}
\includegraphics[width=0.45\textwidth]{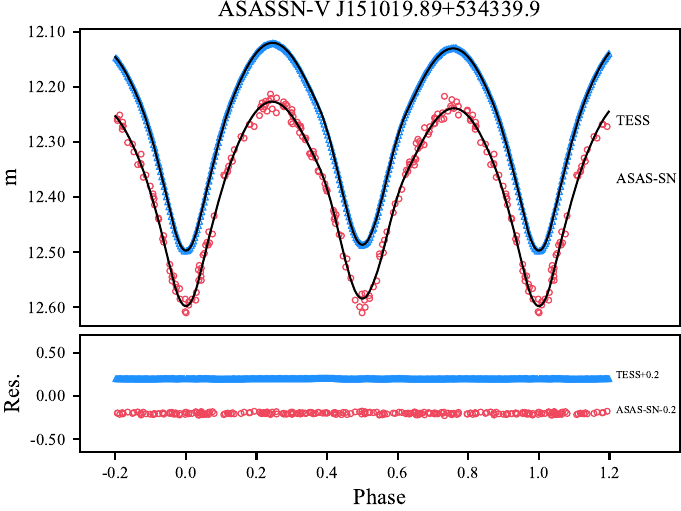}
\caption{The RV and LC fittings of two targets (ASASSN-V J045519.63+451421.6 and ASASSN-V J151019.89+534339.9) are shown as examples.}
\label{fitting}
\end{figure*}

To determine the physical parameters of these systems, we used the 2015 version of the Wilson-Devinney (W-D) code \citep{1971ApJ...166..605W,1979ApJ...234.1054W,1990ApJ...356..613W}.
We downloaded the TESS \citep{2015JATIS...1a4003R} data from the Mikulski Archive for Space Telescopes \citep{2020RNAAS...4..201C},
and found that 138 systems have TESS observations. We then performed simultaneous modeling of ASAS-SN and TESS LCs combined with LAMOST RVs for the 138 contact binaries. For the remaining systems, we analyzed ASAS-SN LCs together with LAMOST RVs. Before modeling, we determine the effective temperature of the primary component. For systems with available temperature measurements from both medium-resolution and low-resolution LAMOST spectra, we adopted the averaged value. For systems with only temperature measurement available from either medium-resolution or low-resolution LAMOST spectra, we directly adopted that value. For systems lacking spectroscopic temperature measurements, we estimated the effective temperature of the primary component from de-reddened B–V, g–r, and J–K color indices and took the mean. The gravity-darkening coefficients and the bolometric albedos were assigned according to \cite{1967ZA.....65...89L} and \cite{1969AcA....19..245R}. For components of contact binaries having temperatures below and above 7200K, we adopted gravity-darkening coefficients of 0.32 and 1.0 \citep{1924MNRAS..84..665V,1967ZA.....65...89L} and bolometric albedos of 0.5 and 1.0 \citep{1969AcA....19..245R}, respectively.
Limb-darkening coefficients were interpolated from the 2019 Van Hamme table\footnote{\url{https://faculty.fiu.edu/~vanhamme/lcdc2015/}} \citep{1993AJ....106.2096V} assuming a logarithmic law (ld = –2). Mode 3 (overcontact configuration) was used. The adjustable parameters during the modeling were mass ratio, orbital inclination, semi-major axis, systemic RV, the effective temperature of the secondary component, the luminosity and dimensionless potential of the primary component. Star spots were introduced on the primary or secondary component to reproduce asymmetric LCs; spot latitude was fixed at $90^\circ$. Physical parameters and best-fit curves were obtained for all 146 targets and are listed in Table \ref{tab:Ph}. We should note that third light ($L_3$) was only included as a free parameter when the synthetic LCs without third light showed a systematically larger amplitude than the observed ones, as an unresolved third light contribution dilutes the observed eclipse depth. When significant third light was detected, it may originate from a physical companion star, a background star within the photometric aperture, or contamination from nearby unresolved stars—an issue particularly relevant for TESS data due to its large pixel size ($~21^{\prime\prime}$ per pixel). The derived third light ratios and their uncertainties are also listed in Table \ref{tab:Ph}. The RV and LC fits of two targets (ASASSN-V J045519.63+451421.6 and ASASSN-V J151019.89+534339.9) are shown as examples in Figure \ref{fitting}.

\begin{table}
\small
\centering
\caption{The physical parameters of the 146 systems} 
\begin{tabular}{llll}
\hline
Number & Column & Units & Explanation                             \\\hline
1      &  Name  &       & ASAS-SN identifier                      \\
2      & Period & day   & Orbital period                          \\
3      & $T_1$  & K     & Temperature of the primary component    \\
4      & $T_2$  & K     & Temperature of the secondary component  \\
5      & $e_{T_2}$& K     & Uncertainty in $T_2$                    \\
6      & q      &       & Mass ratio of the two components        \\
7      & $e_q$  &       & Uncertainty in q                        \\
8      & i      & deg   & Orbital inclination                     \\
9      & $e_i$  & deg   & Uncertainty in i                        \\
10     & a      & $R_\odot$  & Semi-major axis                    \\
11     &$e_a$& $R_\odot$  & Uncertainty in a                      \\
12     &$V_{\gamma}$ & km/s& RV of the system                     \\
13     &$e_{V_{\gamma}}$ & km/s& Uncertainty in $V_{\gamma}$        \\
14     & f     &   & Fillout factor$^*$                               \\
15     & $e_f$     &   & Uncertainty in f                         \\
16     &$L_1/L_V$&     & Light ratio in V band                    \\
17     &$e_{L_1/L_V}$&    & Uncertainty in $L_1/L_V$               \\
18     &$L_1/L_T$&     & Light ratio in Tess band                 \\
19     &$e_{L_1/L_T}$&    & Uncertainty in $L_1/L_T$                \\
20     &$L_3/L_V$&     & Third light ratio in V band              \\
21     &$e_{L_3/L_V}$&    & Uncertainty in $L_3/L_V$                \\
22     &$L_3/L_T$&     & Third light ratio in Tess band           \\
23     &$e_{L_3/L_T}$&    & Uncertainty in $L_3/L_T$                \\
24     &$M_1$&$M_\odot$     & Primary mass                        \\
25     &$e_{M_1}$&$M_\odot$    & Uncertainty in $M_1$              \\
26     &$M_2$&$M_\odot$     & Secondary mass                      \\
27     &$e_{M_2}$&$M_\odot$    & Uncertainty in $M_2$              \\
28     &$R_1$&$R_\odot$     & Primary radius                      \\
29     &$e_{R_1}$&$R_\odot$    & Uncertainty in $R_1$              \\
30     &$R_2$&$R_\odot$     & Secondary radius                    \\
31     &$e_{R_2}$&$R_\odot$    & Uncertainty in $R_2$              \\
32     &$L_1$&$L_\odot$     & Primary Luminosity                  \\
33     &$e_{L_1}$&$L_\odot$    & Uncertainty in $L_1$              \\
34     &$L_2$&$L_\odot$     & Secondary Luminosity                \\
35     &$e_{L_2}$&$L_\odot$    & Uncertainty in $L_2$              \\
36     &Longitude$_1$&    & Spot Longitude of star 1  \\
37     &$r_{s{_1}}$&    & Spot Angular radius of star 1  \\
38     &$T_{s{_1}}$&    & Spot temperature ratio of star 1  \\
39     &Longitude$_2$&    & Spot Longitude of star 2  \\
40     &$r_{s{_2}}$&    & Spot Angular radius of star 2  \\
41     &$T_{s{_2}}$&    & Spot temperature ratio of star 2  \\
\hline
\end{tabular}
\label{tab:Ph}
\\\raggedright $^*$ The definition of fillout factor is $f=$ $\Omega_{in}-\Omega \over \Omega_{in}-\Omega_{out}$, where $\Omega_{in}$ is the potential of the inner critical Roche lobe, $\Omega_{out}$ is the potential of the outer critical Roche lobe, and $\Omega$ is the potential of the two component. 
\end{table}

\section{Discussions and Conclusions}
RVs are crucial for determining the physical parameters of contact binaries, particularly in partial eclipsing systems. This paper derived the RVs from the LAMOST MRS DR9 data for ASAS-SN contact binaries by applying a selection threshold. Based on the analysis of RVs and TESS and ASAS-SN LCs, reliable physical parameters of 146 contact binaries were obtained.

\subsection{Physical parameters comparison with previous studies} \label{sec:physical parameters}
We conducted a historical investigation and identified 10 targets with simultaneous analyzes of RV curves and LCs. The determined physical parameters, including the mass ratio ($q$), orbital inclination ($i$), and fillout factor ($f$), are presented in Table \ref{tab:comparison} alongside historical results for comparison. We found that all targets exhibit a high degree of agreement in their mass ratios (the mass ratio of ASASSN-V J032814.89+300118.4 is inversely proportional to the result reported in the literature, because \cite{2022AJ....163..235L} used the secondary minimum for phase calculation) and orbital inclinations, except ASASSN-V J230344.01+361523.1. For ASASSN-V J230344.01+361523.1, our derived mass ratio ($q=3.43$) differs significantly from that reported by \cite{2025ApJ...979...69W} ($q=2.520$). This discrepancy is likely due to the different number and phase coverage of RV data points: our analysis uses six RV measurements covering nearly one-third of the orbital phase, while Wang et al. (2025) used only three RV points with half the phase coverage. Some targets exhibit significant variation in their fill-out factors, which arises from the use of different LCs.

\begin{table}
\scriptsize
\centering
\caption{Physical parameter comparison between our study and references} 
\begin{tabular}{llllllll}
\hline
Star &                        \multicolumn{2}{c}{$q$}	&	\multicolumn{2}{c}{$i$}	&	\multicolumn{2}{c}{$f$}		&Reference\\
&This paper	    &  Ref. 	        &This paper	  &Ref. 	      &This paper	    &Ref.& \\	\hline 
ASASSN-V J013220.52+551219.7	&$1.454\pm0.005$&	$1.48\pm0.02$	  &$84.5\pm0.0$	&$83.5\pm0.1$	&$14.0\pm1.1\%$	&$12.8\pm4.4\%$	&  \cite{2025AJ....169..139X}\\
ASASSN-V J015112.60+434907.2	&$3.970\pm0.019$&	$3.937\pm0.011$	&$89.5\pm0.5$	&$86.1\pm0.5$	&$24.2\pm4.4\%$	&$12.6\%$	      &  \cite{2015NewA...36..100G}\\
ASASSN-V J032814.89+300118.4	&$2.475\pm0.008$&	$0.402\pm0.001$	&$33.9\pm0.2$	&$33.6\pm0.2$	&$88.8\pm1.2\%$	&$87.0\pm1.6\%$	&  \cite{2022AJ....163..235L}\\
ASASSN-V J034813.48+221850.8	&$0.433\pm0.001$&	$0.439\pm0.001$	&$85.1\pm0.1$	&$84.5\pm0.3$	&$31.3\pm3.5\%$	&$9.3\pm1.1\%$	&  \cite{2015NewA...34..262H}\\
ASASSN-V J085551.39+200340.0	&$1.819\pm0.014$&	$1.76\pm0.05$	  &$73.0\pm0.1$	&$73.6\pm0.1$	&$14.3\pm3.3\%$	&$8.9\pm1.3\%$	&  \cite{2020PASJ...72...73L}\\
ASASSN-V J104212.98+384912.4	&$2.896\pm0.015$&	$2.86\pm0.05$	  &$63.2\pm0.1$	&$62.2\pm0.3$	&$14.1\pm3.1\%$	&$25.7\pm10.7\%$&  \cite{2025AJ....169..139X}\\
ASASSN-V J225840.43+343745.9	&$0.209\pm0.006$&	$0.190\pm0.001$	&$74.9\pm0.5$	&$69.5\pm0.1$	&$43.5\pm8.3\%$	&$22.2\pm1.7\%$	&  \cite{2025ApJ...979...69W}\\
ASASSN-V J230252.69+342300.6	&$0.387\pm0.004$&	$0.373\pm0.003$	&$57.8\pm0.1$	&$55.4\pm0.1$	&$70.1\pm1.9\%$	&$54.2\pm2.7\%$	&  \cite{2025ApJ...979...69W}\\
ASASSN-V J230344.01+361523.1	&$3.434\pm0.016$&	$2.520\pm0.019$	&$66.6\pm0.0$	&$66.7\pm0.1$	&$12.0\pm3.5\%$	&$9.3\pm4.1\%$	&  \cite{2025ApJ...979...69W}\\
ASASSN-V J230813.13+330303.3	&$2.309\pm0.004$&	$2.232\pm0.007$	&$73.2\pm0.1$	&$70.1\pm0.1$	&$19.3\pm2.1\%$	&$11.3\pm1.7\%$	&  \cite{2025ApJ...979...69W}\\
\hline
\end{tabular}
\label{tab:comparison}
\end{table}

\subsection{The evolutionary status}
\begin{figure*}
\centering
\includegraphics[width=0.48\textwidth]{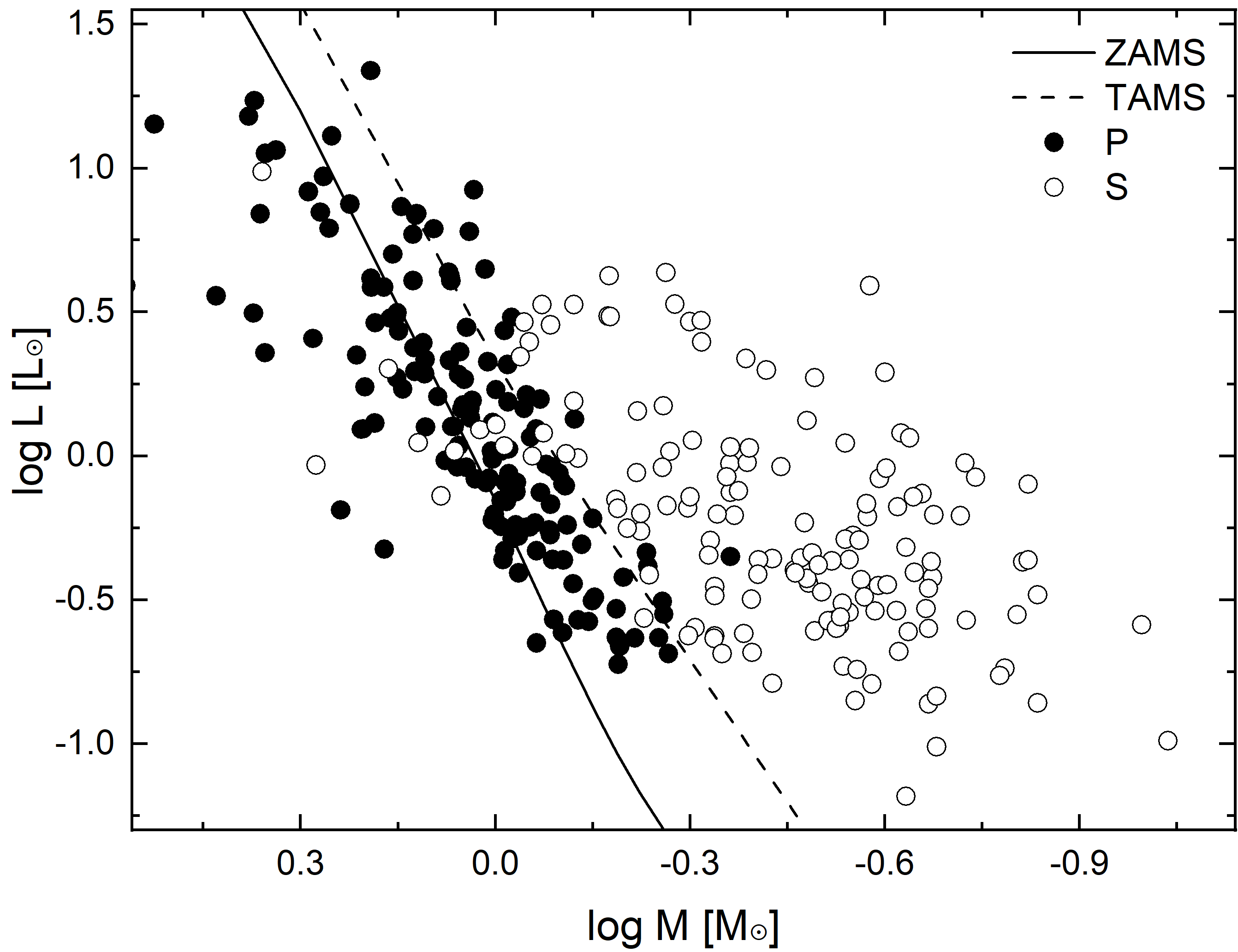}
\includegraphics[width=0.48\textwidth]{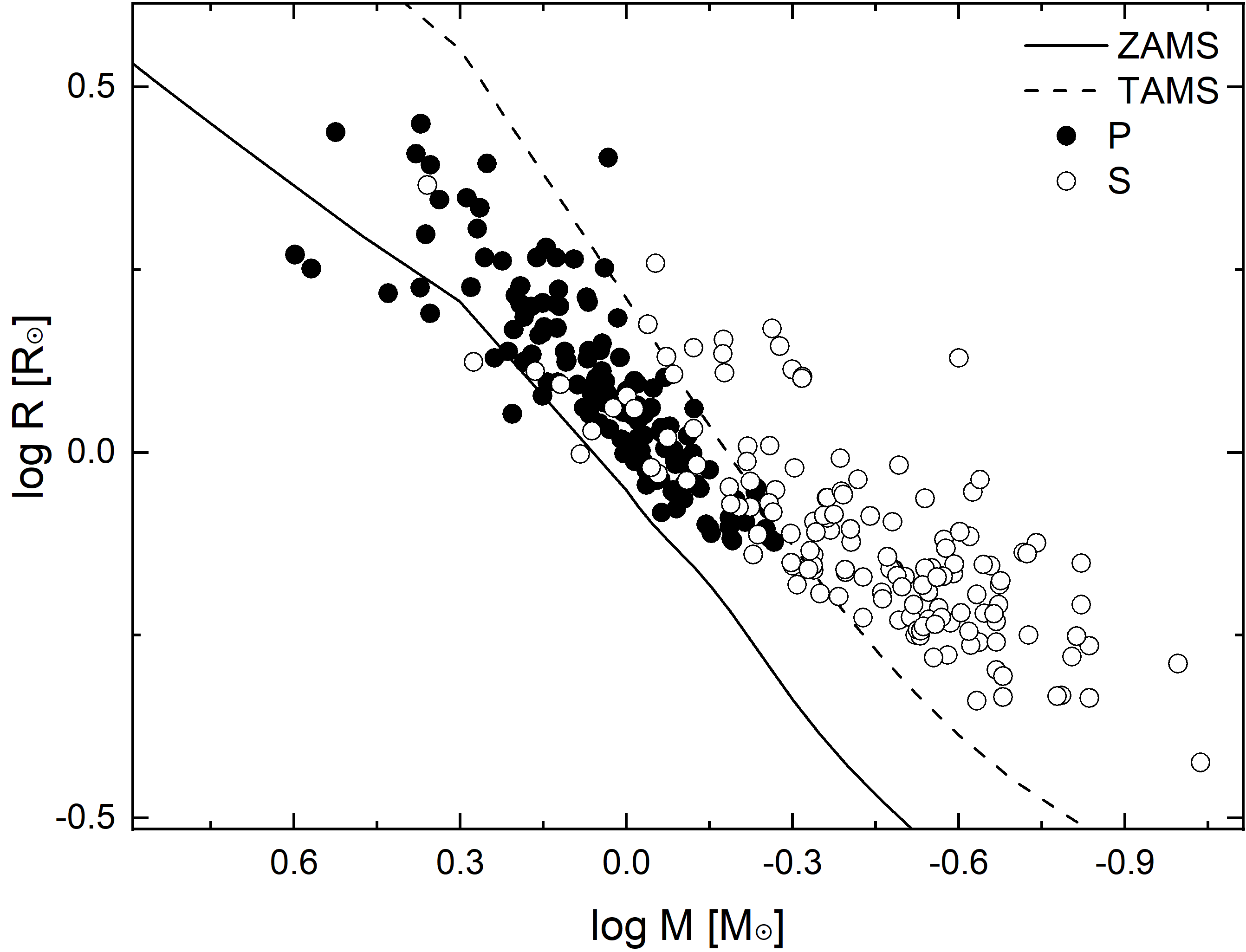}
\includegraphics[width=0.48\textwidth]{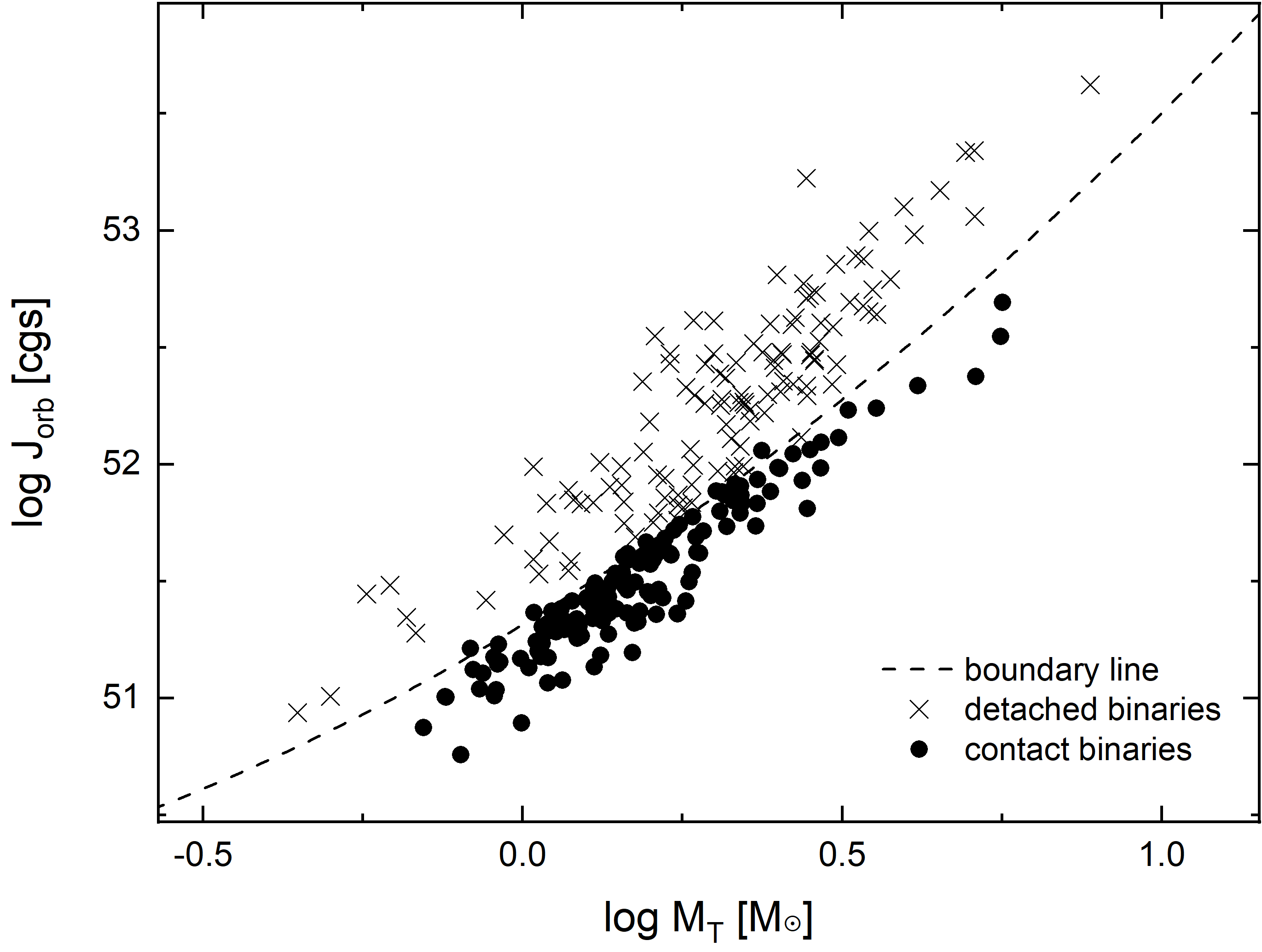}
\caption{The M-L and M-R distributions of the 146 targets are shown in the upper ones. ZAMS and TAMS described by solid and dashed lines are also displayed in this figure. The total mass versus $J_{orb}$ is plotted in the lower one. The dashed line is boundary line between detached and contact binaries derived by \cite{2006MNRAS.373.1483E} and detached binaries are also from \cite{2006MNRAS.373.1483E}.}
\label{MLMR}
\end{figure*}

To elucidate the evolutionary status of our 146 targets, we present their mass-luminosity (M-L) and mass-radius (M-R) distributions in Figure \ref{MLMR}. Based on the binary star evolution code of \cite{2002MNRAS.329..897H}, the zero-age main sequence (ZAMS) and terminal-age main sequence (TAMS) are plotted as solid and dashed lines, respectively. "P" and "S" refer to the more massive and less massive components, respectively. We find that most more massive components are located between the ZAMS and TAMS, whereas the majority of less massive components inhabit the region above the TAMS. This finding aligns with several previous studies (e.g., \citealt{2015AJ....150...69Y,2021AJ....162...13L,2025MNRAS.541.3401Z,2025AJ....170..214P}). The position of the secondaries is generally attributed to mass and energy transfer from the primaries, which makes them appear over-sized and over-luminous relative to single main-sequence stars \citep{2005ApJ...629.1055Y}. However, as demonstrated by \cite{2013MNRAS.430.2029Y}, contact binaries typically experience a mass ratio reversal during their formation and evolution: the current primary was originally the less massive component, while the current secondary was originally the more massive one. The originally more massive star evolved faster, filled its Roche lobe first, and transferred mass to its companion, leading to the mass ratio reversal. Consequently, the current primary has been rejuvenated by mass accretion, appearing less evolved and remaining close to the ZAMS. In contrast, the current secondary has lost a significant fraction of its original mass and is now in a thermally expanded state, which provides an alternative explanation for its over-sized and over-luminous appearance—even without invoking additional mass transfer from the current primary. Therefore, the location of the primaries between ZAMS and TAMS does not necessarily indicate significant nuclear evolution; rather, it reflects the combined effects of mass accretion, thermal relaxation, and energy transfer. For the secondaries, their position above the TAMS could be due to either (2) mass and energy gain from the primaries \citep{2005ApJ...629.1055Y}; or (1) mass loss and thermal expansion following mass ratio reversal \citep{2013MNRAS.430.2029Y}. Determining the precise evolutionary pathways that lead to the observed positions of both components would require detailed modeling of their evolutionary history, which is beyond the scope of this study.

To investigate the formation mechanism of the contact binary, we computed the orbital angular momentum ($J_{orb}$) using the following equation from \citet{2013AJ....146..157C},
\begin{equation}
\label{eq:J_o}
J_{orb} = 1.24 \times 10^{52} \times M_{total}^{\frac{5}{3}} \times P^{\frac{1}{3}} \times {\frac{q}{(1+q)^2}}.
\end{equation}
The total mass versus $J_{orb}$ is plotted in Figure \ref{MLMR}. The dashed line, adopted from \cite{2006MNRAS.373.1483E}, marks the boundary between detached and contact binaries. The data points for detached binaries in the figure are also sourced from the same study. For a given total mass, contact binaries possess systematically smaller orbital angular momentum than detached binaries. This observed distribution is consistent with the prevailing theory that contact binaries form from detached binaries through AML via magnetic stellar wind (e.g., \citealt{1988ASIC..241..345G,1994ASPC...56..228B,2017RAA....17...87Q}). We note that the lower $J_{orb}$ values are inherently linked to the shorter orbital separations and periods defining the contact configuration. Thus, while Figure \ref{MLMR} aligns with the AML formation scenario, it represents the systems' current dynamical states rather than direct proof of their evolutionary history.

\subsection{Special targets}
Contact binaries with small mass ratios ($q < 0.25$), particularly those with extremely low mass ratios ($q < 0.15$), are regarded as progenitors of stellar mergers (e.g., \citealt{2006Ap&SS.304...25Q,2011A&A...528A.114T,2016RAA....16...68Z,2022AJ....164..202L,2024A&A...692L...4L}). Among our analyzed 146 systems, there are 38 low mass ratio contact binaries, and 11 extremely low mass ratio ones. Notably, the system ASASSN-V J111451.48+005038.6 exhibits a mass ratio of 0.113 and a fill-out factor of 98.3$\pm3.7$\%. To our knowledge, this target represents the highest fill-out factor reported for a contact binary to date. The combination of an extremely low mass ratio and the highest fill-out factor makes this system a prominent candidate for merger in the near future.

High mass ratio ($q > 0.72$, H-type, \citealt{2004A&A...426.1001C}) contact binaries provide critical insight into close binary evolution, especially the transition from near-contact to contact phases \citep{2023AJ....166..200C,2025AJ....169...85X,2025AJ....170..167L}. Our sample contains 11 H-type contact binaries, ten of which are in shallow contact. Notably, ASASSN-V J093921.74+390452.6, with a mass ratio of $0.993\pm0.009$, represents the contact binary with the highest spectroscopically determined mass ratio known to date. Its low fillout factor of only 1.8$\pm2.3$\% and high mass ratio indicate that the system achieved contact very recently.

\subsection{Several correlations between the physical parameters}
\begin{figure}\centering
\includegraphics[width=0.45\textwidth]{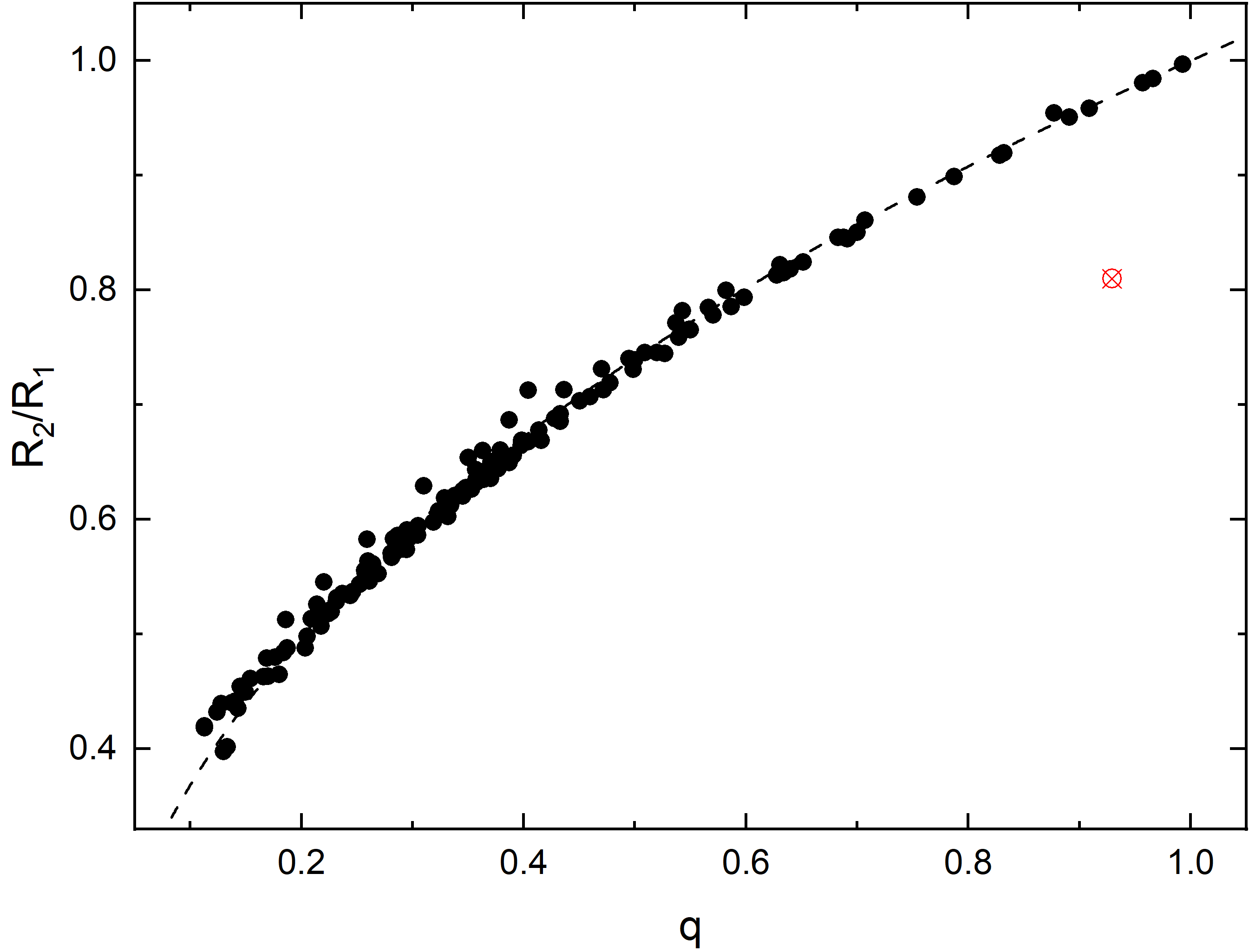}
\includegraphics[width=0.45\textwidth]{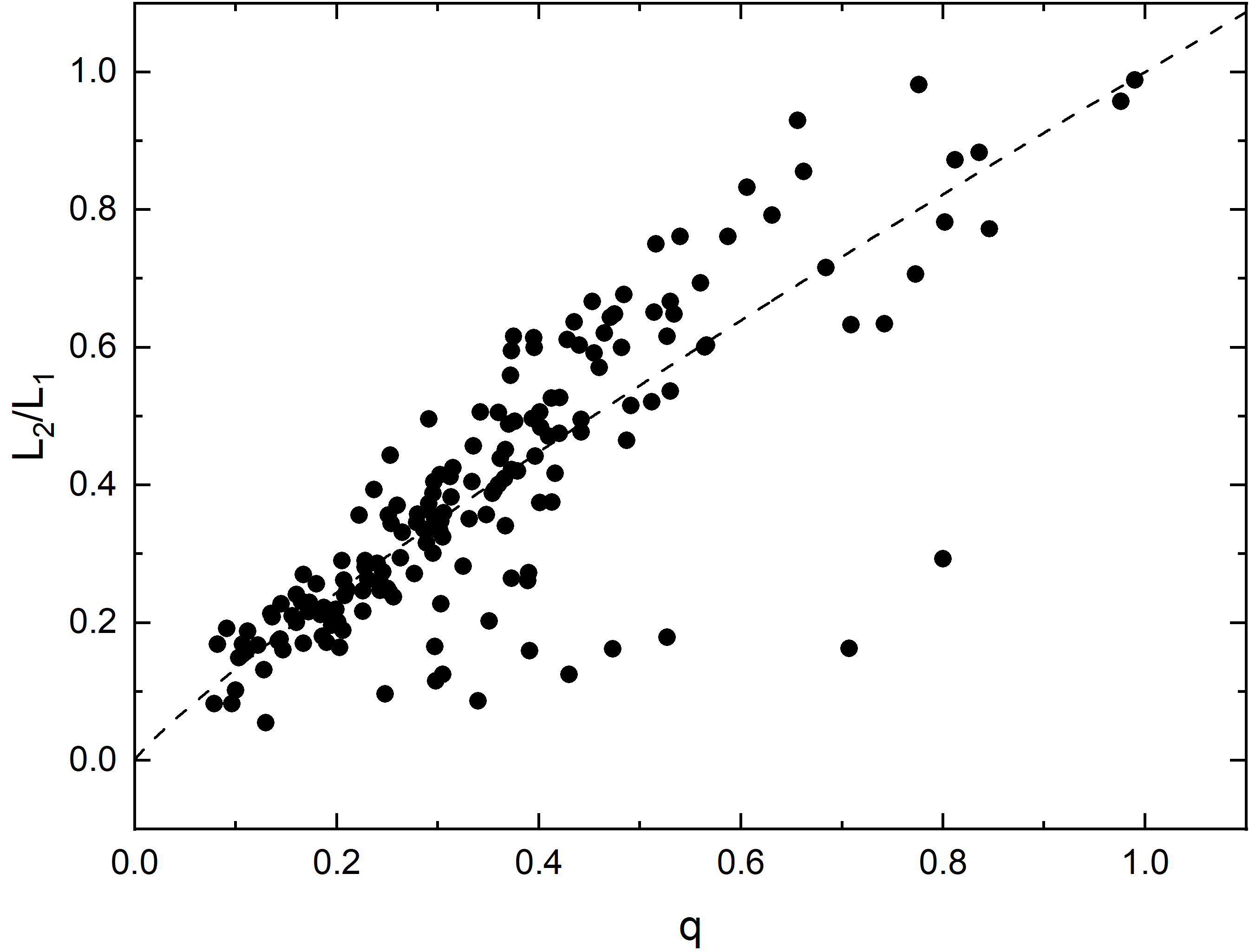}
\includegraphics[width=0.45\textwidth]{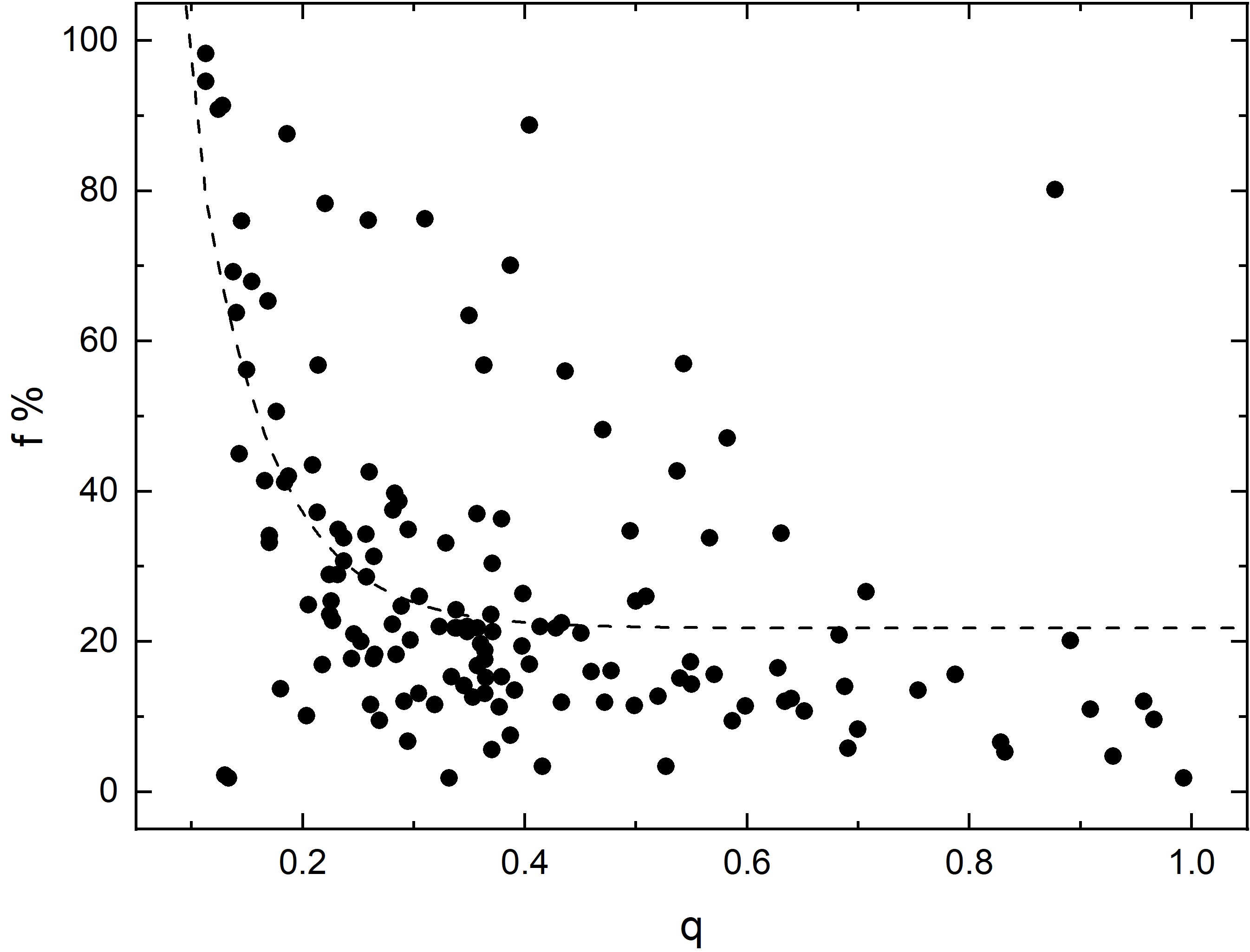}
\includegraphics[width=0.455\textwidth]{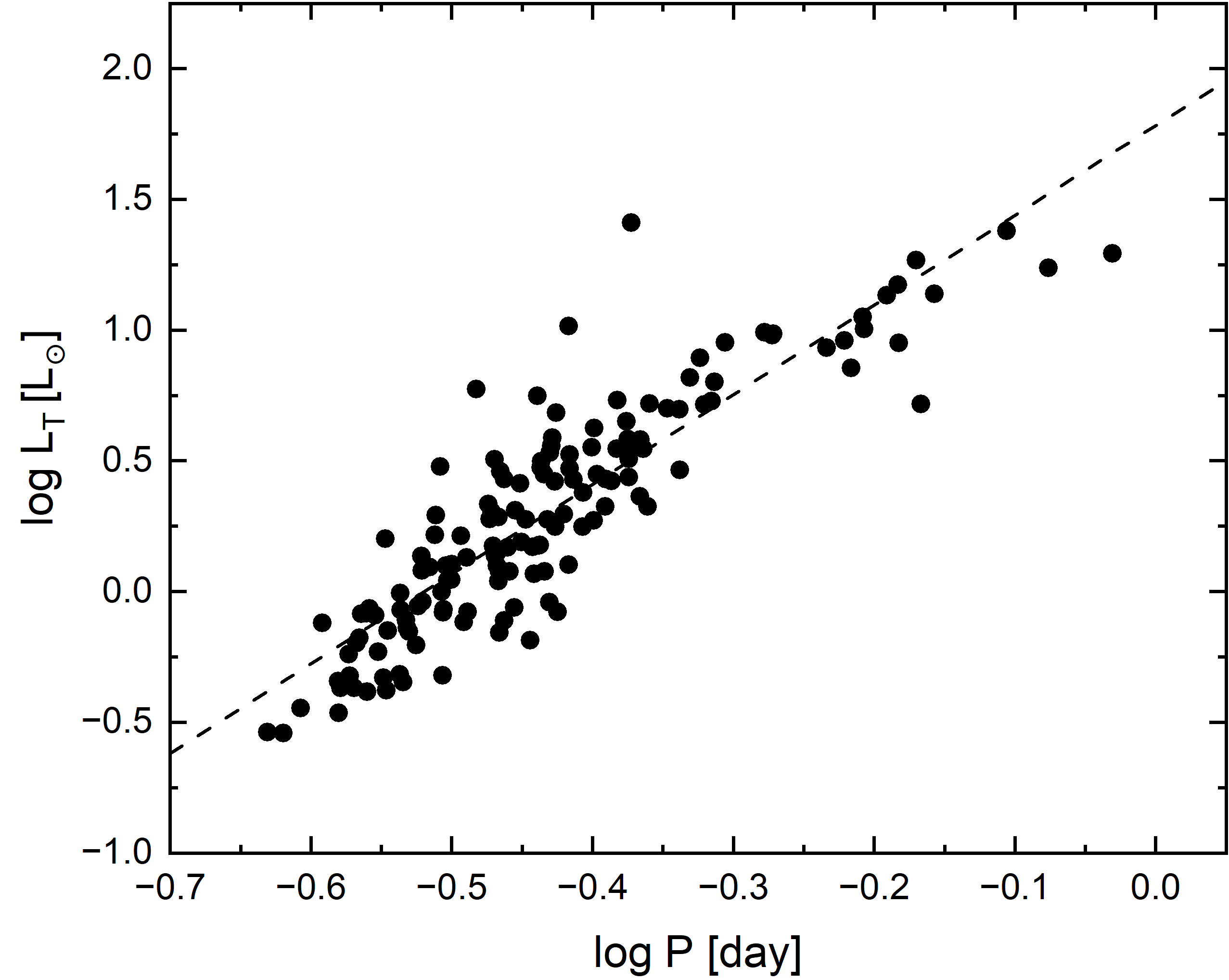}
\includegraphics[width=0.45\textwidth]{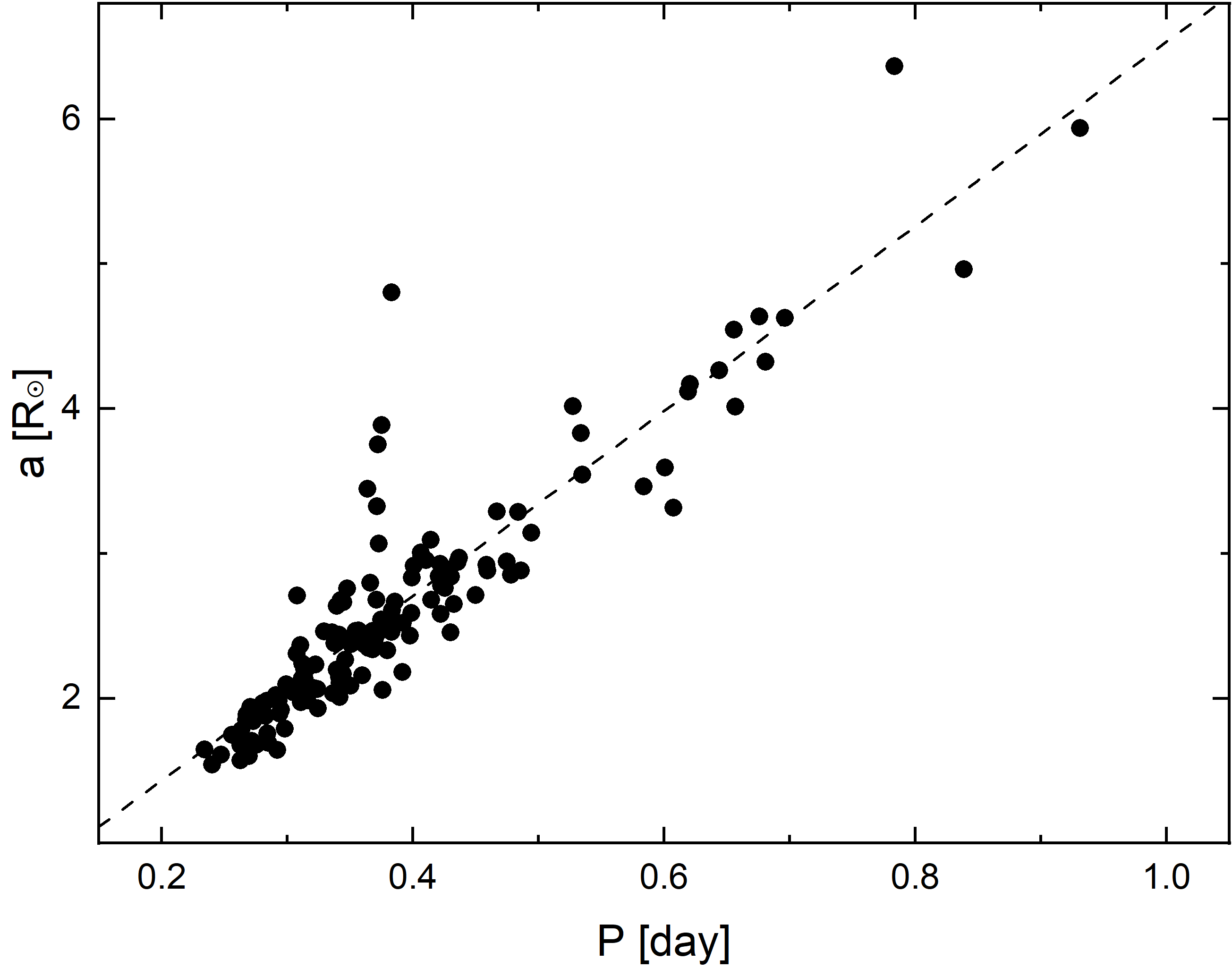}
\includegraphics[width=0.475\textwidth]{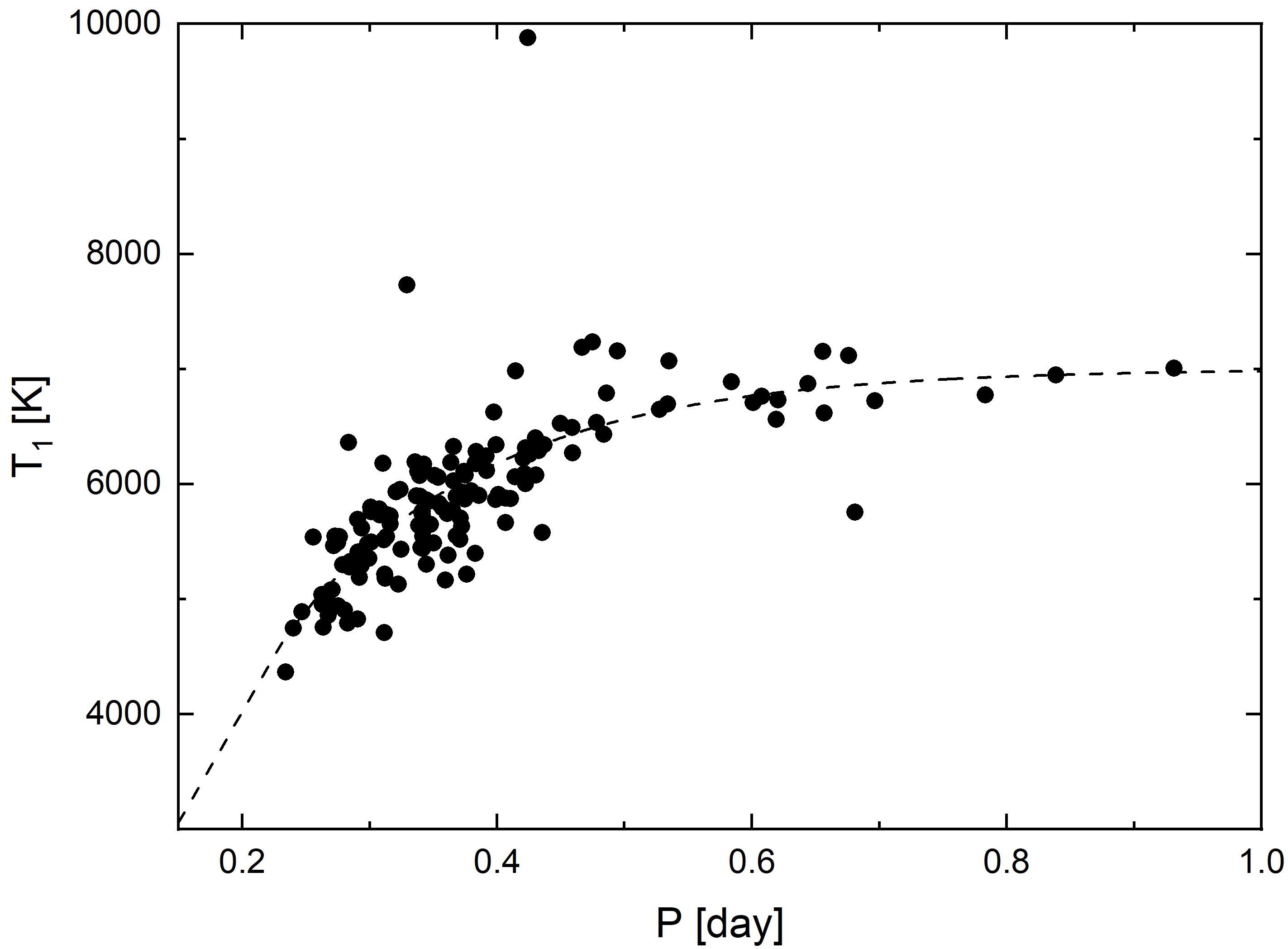}
\caption{The relations of $q-R_2/R_1$, $q-L_2/L_1$, $q-f$, $\log P-\log L_T$, $P-a$, and $P-T_1$ for the 146 contact binaries. The red circle in the $q-R_2/R_1$ figure is deleted during the fitting.}
\label{Relation}
\end{figure}

Based on the sample of 146 contact binaries, several correlations between their physical parameters were investigated. The relationships of $q-R_2/R_1$, $q-L_2/L_1$, $q-f$, $\log P-\log L_T$, $P-a$, and $P-T_1$ are presented in Figure \ref{Relation}, where $q$, $R$, $L$, $f$, $P$, $L_T$, $a$, and $T_1$ denote the mass ratio, radius, luminosity, contact degree, orbital period, the total luminosity of a binary, semi-major axis, the temperature of the primary component, respectively.
Our analysis yielded the following results:

(1) For the majority of the sample, the radius ratio falls within the range of 0.3 to 1.0. We find that the observed radius ratio correlates with the mass ratio, which can be empirically described by the following relation:
\begin{eqnarray}
R_2/R_1=q^{0.434(\pm0.001)}.
\end{eqnarray}
It should be noted that in the Roche model, the radius ratio is a function of the mass ratio and the filling factor, and this determined relation is consistent with the findings of \cite{1941ApJ....93..133K}. Therefore, the empirical relation presented here reflects the statistical trend within our specific sample, rather than a new fundamental physical law. Only one binary exhibits a significant deviation from this general trend (the red circle in Figure \ref{Relation}).

(2) A strong exponential correlation between the luminosity ratio and mass ratio is obtained,
\begin{eqnarray}
L_2/L_1=q^{0.877(\pm0.023)}.
\end{eqnarray}
This relation is also a reflection of the Roche model and is very close to the result reported by \cite{1968ApJ...151.1123L}.

(3) Deep contact systems ($f>50\%$) account for less than one-fifth of the sample. A clear trend is observed wherein the fill-out factor increases as the mass ratio decreases. This behavior is described by the following equation,
\begin{eqnarray}
f=21.768(\pm2.308)+312.722(\pm164.917)e^{-15.040(\pm3.862)q}.
\end{eqnarray}
It is important to note that the observed relative scarcity of deep-contact systems may be influenced by selection effects inherent to the LAMOST/ASAS-SN survey. A statistically robust interpretation regarding the true population frequency of deep-contact binaries would require a thorough assessment of sample completeness, which is beyond the scope of this parameter analysis.

(4) Consistent with the known period–luminosity relation (e.g., \citealt{2017AJ....154..125M,2018ApJ...859..140C}), the total luminosity of the systems increases with orbital period, as given by,
\begin{eqnarray}
\log L_T=3.430(\pm0.151)+1.782(\pm0.067)\log P.
\end{eqnarray}
The strong correlation provides further evidence supporting the validity of the period–luminosity relation for contact binaries, and highlights the utility of orbital period as a robust indicator of system luminosity in such objects.

(5) The semi-major axis for most systems lies in the range of 1.5–6.0 $R_\odot$. A well-defined linear relation between the orbital period and semi-major axis is derived,
\begin{eqnarray}
a=0.158(\pm0.098)+6.374(\pm0.243)P.
\end{eqnarray}
This empirical relation can serve as a convenient tool for estimating the absolute parameters of contact binary systems. It’s physical basis can be understood through Kepler’s Third Law ($a \propto M^{1/3} P^{2/3}$) in conjunction with the approximate mass-radius relation ($R \propto M$) for low-mass main-sequence stars. For contact binaries where the component stars are in physical contact ($R \propto a$), these relations combine to yield an approximate linear relation between the semi-major axis and orbital period $(a \propto P$). 

(6) The primary component temperatures for most binaries range from 4000 K to 8000 K. A noticeable increasing trend is seen in the primary temperature with orbital period, as modeled by,
\begin{eqnarray}
T_1=6999.633(\pm195.660)-10115.249(\pm3082.410)e^{-6.285(\pm1.357)P}.
\end{eqnarray}
This behavior aligns with the established period–color relation of contact binaries (e.g., \citealt{1998AJ....116.2998R}).

In summary, this study demonstrates the successful use of the LAMOST MRS survey to determine reliable RV curves.  By combining these with photometric data from ASAS-SN and TESS, we derived physical parameters for 146 systems, substantially expanding the catalog of contact binaries with precise photometric and spectroscopic solutions, providing a valuable resource for understanding their physical properties and evolutionary status.
The evolutionary status investigation reveals that the more massive primaries are typically less-evolved main-sequence stars, while the lower-mass secondaries are over-sized and over-luminous due to mass and energy transfer.
Furthermore, the orbital angular momentum distribution relative to total mass supports the formation scenario in which contact binaries evolve from detached binaries via AML.
Significantly, we identified systems in extreme evolutionary states, including low mass ratio binaries that are potential stellar merger candidates, and high mass ratio systems in shallow contact that may have only recently formed. These objects offer unique insights into both the onset and end stages of contact binary evolution.
Finally, we established several empirical relations—between mass ratio and radius/luminosity ratios, fill-out factor, as well as orbital period with total luminosity, semi-major axis, and primary temperature. These relations provide powerful tools for system characterization, absolute parameter determination, and distance estimation, while also offering crucial constraints for theoretical models of contact binary evolution.

\begin{acknowledgments}
We sincerely thank the reviewer for the valuable comments and constructive suggestions, which have significantly improved the quality of our manuscript. This work is supported by National Natural Science Foundation of China (NSFC) (No. 12273018), and by the Taishan Scholars Young Expert Program of Shandong Province, and by the Qilu Young Researcher Project of Shandong University, and by the Young Data Scientist Project of the National Astronomical Data Center. The calculations in this work were carried out at Supercomputing Center of Shandong University, Weihai.

The spectral data were provided by Guoshoujing Telescope (the Large Sky Area Multi-Object Fiber Spectroscopic Telescope; LAMOST), which is a National Major Scientific Project built by the Chinese Academy of Sciences. Funding for the project has been provided by the National Development and Reform Commission. LAMOST is operated and managed by the National Astronomical Observatories, Chinese Academy of Sciences.

This paper makes use of data from ASAS-SN. ASAS-SN is funded
in part by the Gordon and Betty Moore Foundation through grant numbers GBMF5490 and GBMF10501 to the Ohio State University, and also funded in part by the Alfred P. Sloan Foundation grant number G-2021-14192. 

This paper includes data collected by TESS. Funding for TESS is provided by the NASA’s Science Mission Directorate. TESS data in this paper were obtained from the Mikulski Archive for Space Telescopes (MAST) at the Space Telescope Science Institute.

\end{acknowledgments}

\bibliography{sample631}{}

@INPROCEEDINGS{1994ASPC...56..228B,
       author = {{Bradstreet}, D.~H. and {Guinan}, E.~F.},
        title = "{Stellar Mergers and Acquisitions: The Formation and Evolution of W Ursae Majoris Binaries}",
    booktitle = {Interacting Binary Stars},
         year = 1994,
       editor = {{Shafter}, A.~W.},
       series = {Astronomical Society of the Pacific Conference Series},
       volume = {56},
        month = jan,
        pages = {228},
       adsurl = {https://ui.adsabs.harvard.edu/abs/1994ASPC...56..228B}
}

@ARTICLE{2019A&A...628A..78R,
       author = {{Riener}, M. and {Kainulainen}, J. and {Henshaw}, J.~D. and {Orkisz}, J.~H. and {Murray}, C.~E. and {Beuther}, H.},
        title = "{GAUSSPY+: A fully automated Gaussian decomposition package for emission line spectra}",
      journal = {\aap},
         year = 2019,
        month = aug,
       volume = {628},
          eid = {A78},
        pages = {A78},
          doi = {10.1051/0004-6361/201935519},
archivePrefix = {arXiv},
       eprint = {1906.10506},
 primaryClass = {astro-ph.IM},
       adsurl = {https://ui.adsabs.harvard.edu/abs/2019A&A...628A..78R}
}

@ARTICLE{https://doi.org/10.17909/fwdt-2x66,
  doi = {10.17909/FWDT-2X66},
  url = {http://archive.stsci.edu/doi/resolve/resolve.html?doi=10.17909/fwdt-2x66},
  author = {{TESS Team}},
  title = {TESS Input Catalog and Candidate Target List},
  publisher = {STScI/MAST},
  year = {2018}
}

@ARTICLE{2024A&A...692L...4L,
       author = {{Li}, Kai and {Gao}, Xiang and {Guo}, Di-Fu and {Gao}, Dong-Yang and {Chen}, Xu and {Wang}, Li-Heng and {Xin}, Yu-Xin and {Han}, Yu-Xin and {Kim}, Chun-Hwey and {Jeong}, Min-Ji},
        title = "{Detection of the lowest mass ratio contact binary in the universe: TYC 3801-1529-1}",
      journal = {\aap},
         year = 2024,
        month = dec,
       volume = {692},
          eid = {L4},
        pages = {L4},
          doi = {10.1051/0004-6361/202451947},
archivePrefix = {arXiv},
       eprint = {2411.12132},
 primaryClass = {astro-ph.SR},
       adsurl = {https://ui.adsabs.harvard.edu/abs/2024A&A...692L...4L}
}

@ARTICLE{2022AJ....164..202L,
       author = {{Li}, Kai and {Gao}, Xiang and {Liu}, Xin-Yi and {Gao}, Xing and {Li}, Ling-Zhi and {Chen}, Xu and {Sun}, Guo-You},
        title = "{Extremely Low Mass Ratio Contact Binaries. I. The First Photometric and Spectroscopic Investigations of Ten Systems}",
      journal = {\aj},
         year = 2022,
        month = nov,
       volume = {164},
       number = {5},
          eid = {202},
        pages = {202},


       eprint = {2209.03653},
 primaryClass = {astro-ph.SR}
}

@ARTICLE{2024MNRAS.527..521K,
       author = {{Kovalev}, Mikhail and {Zhou}, Zenghua and {Chen}, Xuefei and {Han}, Zhanwen},
        title = "{Detection of 12 426 SB2 candidates in the LAMOST-MRS, using a binary spectral model}",
      journal = {\mnras},
         year = 2024,
        month = jan,
       volume = {527},
       number = {1},
        pages = {521-530},
          doi = {10.1093/mnras/stad3222},
archivePrefix = {arXiv},
       eprint = {2310.11673},
 primaryClass = {astro-ph.SR},
       adsurl = {https://ui.adsabs.harvard.edu/abs/2024MNRAS.527..521K}
}

@ARTICLE{2013A&A...553A...6H,
       author = {{Husser}, T. -O. and {Wende-von Berg}, S. and {Dreizler}, S. and {Homeier}, D. and {Reiners}, A. and {Barman}, T. and {Hauschildt}, P.~H.},
        title = "{A new extensive library of PHOENIX stellar atmospheres and synthetic spectra}",
      journal = {\aap},
         year = 2013,
        month = may,
       volume = {553},
          eid = {A6},
        pages = {A6},
          doi = {10.1051/0004-6361/201219058},
archivePrefix = {arXiv},
       eprint = {1303.5632},
 primaryClass = {astro-ph.SR},
       adsurl = {https://ui.adsabs.harvard.edu/abs/2013A&A...553A...6H}
}

@ARTICLE{2025ApJS..276...11L,
       author = {{Li}, Shan-shan and {Li}, Chun-qian and {Li}, Chang-hua and {Fan}, Dong-wei and {Xu}, Yun-fei and {Mi}, Lin-ying and {Cui}, Chen-zhou and {Shi}, Jian-rong},
        title = "{Mining Double-line Spectroscopic Candidates in the LAMOST Medium-resolution Spectroscopic Survey Using a Human{\textendash}AI Hybrid Method}",
      journal = {\apjs},
         year = 2025,
        month = jan,
       volume = {276},
       number = {1},
          eid = {11},
        pages = {11},
          doi = {10.3847/1538-4365/ad9010},
archivePrefix = {arXiv},
       eprint = {2411.14714},
 primaryClass = {astro-ph.IM},
       adsurl = {https://ui.adsabs.harvard.edu/abs/2025ApJS..276...11L}
}

@ARTICLE{2022ApJS..258...16P,
       author = {{Pr{\v{s}}a}, Andrej and {Kochoska}, Angela and {Conroy}, Kyle E. and {Eisner}, Nora and {Hey}, Daniel R. and {IJspeert}, Luc and {Kruse}, Ethan and {Fleming}, Scott W. and {Johnston}, Cole and {Kristiansen}, Martti H. and {LaCourse}, Daryll and {Mortensen}, Danielle and {Pepper}, Joshua and {Stassun}, Keivan G. and {Torres}, Guillermo and {Abdul-Masih}, Michael and {Chakraborty}, Joheen and {Gagliano}, Robert and {Guo}, Zhao and {Hambleton}, Kelly and {Hong}, Kyeongsoo and {Jacobs}, Thomas and {Jones}, David and {Kostov}, Veselin and {Lee}, Jae Woo and {Omohundro}, Mark and {Orosz}, Jerome A. and {Page}, Emma J. and {Powell}, Brian P. and {Rappaport}, Saul and {Reed}, Phill and {Schnittman}, Jeremy and {Schwengeler}, Hans Martin and {Shporer}, Avi and {Terentev}, Ivan A. and {Vanderburg}, Andrew and {Welsh}, William F. and {Caldwell}, Douglas A. and {Doty}, John P. and {Jenkins}, Jon M. and {Latham}, David W. and {Ricker}, George R. and {Seager}, Sara and {Schlieder}, Joshua E. and {Shiao}, Bernie and {Vanderspek}, Roland and {Winn}, Joshua N.},
        title = "{TESS Eclipsing Binary Stars. I. Short-cadence Observations of 4584 Eclipsing Binaries in Sectors 1-26}",
      journal = {\apjs},
         year = 2022,
        month = jan,
       volume = {258},
       number = {1},
          eid = {16},
        pages = {16},
          doi = {10.3847/1538-4365/ac324a},
archivePrefix = {arXiv},
       eprint = {2110.13382},
 primaryClass = {astro-ph.SR},
       adsurl = {https://ui.adsabs.harvard.edu/abs/2022ApJS..258...16P}
}

@INPROCEEDINGS{1988ASIC..241..345G,
       author = {{Guinan}, Edward F. and {Bradstreet}, David H.},
        title = "{Kinematic Clues to the Origin and Evolution of Low Mass Contact Binaries}",
    booktitle = {Formation and Evolution of Low Mass Stars},
         year = 1988,
       editor = {{Dupree}, A.~K. and {Lago}, M.~T.~V.~T.},
       series = {NATO Advanced Study Institute (ASI) Series C},
       volume = {241},
        month = jan,
        pages = {345},
       adsurl = {https://ui.adsabs.harvard.edu/abs/1988ASIC..241..345G}
}

@ARTICLE{2003MNRAS.342.1260Q,
       author = {{Qian}, Shengbang},
        title = "{Are overcontact binaries undergoing thermal relaxation oscillation with variable angular momentum loss?}",
      journal = {\mnras},
         year = 2003,
        month = jul,
       volume = {342},
       number = {4},
        pages = {1260-1270},
          doi = {10.1046/j.1365-8711.2003.06627.x},
       adsurl = {https://ui.adsabs.harvard.edu/abs/2003MNRAS.342.1260Q}
}

@ARTICLE{2006Ap&SS.304...25Q,
       author = {{Qian}, Shengbang and {Yang}, Yuangui and {Zhu}, Liying and {He}, Jiajia and {Yuan}, Jingzhao},
        title = "{Photometric Studies of Twelve Deep, Low-mass Ratio Overcontact Binary Systems}",
      journal = {\apss},
         year = 2006,
        month = aug,
       volume = {304},
       number = {1-4},
        pages = {25-28},
          doi = {10.1007/s10509-006-9114-z},
       adsurl = {https://ui.adsabs.harvard.edu/abs/2006Ap&SS.304...25Q}
}

@ARTICLE{2017RAA....17...87Q,
       author = {{Qian}, Sheng-Bang and {He}, Jia-Jia and {Zhang}, Jia and {Zhu}, Li-Ying and {Shi}, Xiang-Dong and {Zhao}, Er-Gang and {Zhou}, Xiao},
        title = "{Physical properties and catalog of EW-type eclipsing binaries observed by LAMOST}",
      journal = {Research in Astronomy and Astrophysics},
         year = 2017,
        month = aug,
       volume = {17},
       number = {8},
          eid = {087},
        pages = {087},
          doi = {10.1088/1674-4527/17/8/87},
archivePrefix = {arXiv},
       eprint = {1705.03996},
 primaryClass = {astro-ph.SR},
       adsurl = {https://ui.adsabs.harvard.edu/abs/2017RAA....17...87Q}
}

@ARTICLE{2005ApJ...629.1055Y,
       author = {{Yakut}, Kadri and {Eggleton}, Peter P.},
        title = "{Evolution of Close Binary Systems}",
      journal = {\apj},
         year = 2005,
        month = aug,
       volume = {629},
       number = {2},
        pages = {1055-1074},
          doi = {10.1086/431300},
       adsurl = {https://ui.adsabs.harvard.edu/abs/2005ApJ...629.1055Y}
}

@ARTICLE{2006AcA....56..347S,
       author = {{Stepien}, K.},
        title = "{The Low-Mass Limit for Total Mass of W UMa-type Binaries}",
      journal = {\actaa},
         year = 2006,
        month = dec,
       volume = {56},
        pages = {347-364},
archivePrefix = {arXiv},
       eprint = {astro-ph/0701529},
 primaryClass = {astro-ph},
       adsurl = {https://ui.adsabs.harvard.edu/abs/2006AcA....56..347S}
}

@ARTICLE{2011A&A...528A.114T,
       author = {{Tylenda}, R. and {Hajduk}, M. and {Kami{\'n}ski}, T. and {Udalski}, A. and {Soszy{\'n}ski}, I. and {Szyma{\'n}ski}, M.~K. and {Kubiak}, M. and {Pietrzy{\'n}ski}, G. and {Poleski}, R. and {Wyrzykowski}, {\L}. and {Ulaczyk}, K.},
        title = "{V1309 Scorpii: merger of a contact binary}",
      journal = {\aap},
         year = 2011,
        month = apr,
       volume = {528},
          eid = {A114},
        pages = {A114},
          doi = {10.1051/0004-6361/201016221},
archivePrefix = {arXiv},
       eprint = {1012.0163},
 primaryClass = {astro-ph.SR},
       adsurl = {https://ui.adsabs.harvard.edu/abs/2011A&A...528A.114T}
}

@ARTICLE{1995ApJ...444L..41R,
       author = {{Rasio}, Frederic A.},
        title = "{The Minimum Mass Ratio of W Ursae Majoris Binaries}",
      journal = {\apjl},
         year = 1995,
        month = may,
       volume = {444},
        pages = {L41},
          doi = {10.1086/187855},
archivePrefix = {arXiv},
       eprint = {astro-ph/9502028},
 primaryClass = {astro-ph},
       adsurl = {https://ui.adsabs.harvard.edu/abs/1995ApJ...444L..41R}
}

@ARTICLE{2006MNRAS.369.2001L,
       author = {{Li}, Lifang and {Zhang}, Fenghui},
        title = "{The dynamical stability of W Ursae Majoris-type systems}",
      journal = {\mnras},
         year = 2006,
        month = jul,
       volume = {369},
       number = {4},
        pages = {2001-2004},
          doi = {10.1111/j.1365-2966.2006.10462.x},
archivePrefix = {arXiv},
       eprint = {0805.4041},
 primaryClass = {astro-ph},
       adsurl = {https://ui.adsabs.harvard.edu/abs/2006MNRAS.369.2001L}
}

@ARTICLE{2021MNRAS.501..229W,
       author = {{Wadhwa}, Surjit S. and {De Horta}, Ain and {Filipovi{\'c}}, Miroslav D. and {Tothill}, N.~F.~H. and {Arbutina}, Bojan and {Petrovi{\'c}}, Jelena and {Djura{\v{s}}evi{\'c}}, Gojko},
        title = "{ZZ Piscis Austrinus (ZZ PsA): a bright red nova progenitor and the instability mass ratio of contact binary stars}",
      journal = {\mnras},
         year = 2021,
        month = jan,
       volume = {501},
       number = {1},
        pages = {229-235},
          doi = {10.1093/mnras/staa3637},
archivePrefix = {arXiv},
       eprint = {2011.09090},
 primaryClass = {astro-ph.SR},
       adsurl = {https://ui.adsabs.harvard.edu/abs/2021MNRAS.501..229W}
}

@ARTICLE{2014ApJ...788...48S,
       author = {{Shappee}, B.~J. and {Prieto}, J.~L. and {Grupe}, D. and {Kochanek}, C.~S. and {Stanek}, K.~Z. and {De Rosa}, G. and {Mathur}, S. and {Zu}, Y. and {Peterson}, B.~M. and {Pogge}, R.~W. and {Komossa}, S. and {Im}, M. and {Jencson}, J. and {Holoien}, T.~W. -S. and {Basu}, U. and {Beacom}, J.~F. and {Szczygie{\l}}, D.~M. and {Brimacombe}, J. and {Adams}, S. and {Campillay}, A. and {Choi}, C. and {Contreras}, C. and {Dietrich}, M. and {Dubberley}, M. and {Elphick}, M. and {Foale}, S. and {Giustini}, M. and {Gonzalez}, C. and {Hawkins}, E. and {Howell}, D.~A. and {Hsiao}, E.~Y. and {Koss}, M. and {Leighly}, K.~M. and {Morrell}, N. and {Mudd}, D. and {Mullins}, D. and {Nugent}, J.~M. and {Parrent}, J. and {Phillips}, M.~M. and {Pojmanski}, G. and {Rosing}, W. and {Ross}, R. and {Sand}, D. and {Terndrup}, D.~M. and {Valenti}, S. and {Walker}, Z. and {Yoon}, Y.},
        title = "{The Man behind the Curtain: X-Rays Drive the UV through NIR Variability in the 2013 Active Galactic Nucleus Outburst in NGC 2617}",
      journal = {\apj},
         year = 2014,
        month = jun,
       volume = {788},
       number = {1},
          eid = {48},
        pages = {48},
          doi = {10.1088/0004-637X/788/1/48},
archivePrefix = {arXiv},
       eprint = {1310.2241},
 primaryClass = {astro-ph.HE},
       adsurl = {https://ui.adsabs.harvard.edu/abs/2014ApJ...788...48S}
}

@ARTICLE{2015NewA...36..100G,
       author = {{G{\"u}rol}, Birol and {Bradstreet}, David H. and {Okan}, Abdullah},
        title = "{Absolute and geometric parameters of the W UMa type contact binary V546 And}",
      journal = {\na},
         year = 2015,
        month = apr,
       volume = {36},
        pages = {100-109},
          doi = {10.1016/j.newast.2014.10.007},
       adsurl = {https://ui.adsabs.harvard.edu/abs/2015NewA...36..100G}
}

@ARTICLE{2022AJ....163..235L,
       author = {{Lei}, Yajuan and {Li}, Guangwei and {Zhou}, Guiping and {Li}, Chunqian},
        title = "{Analysis of Five Double-lined Spectroscopic Eclipsing Binaries Observed with TESS and LAMOST}",
      journal = {\aj},
         year = 2022,
        month = may,
       volume = {163},
       number = {5},
          eid = {235},
        pages = {235},
          doi = {10.3847/1538-3881/ac5aa5},
       adsurl = {https://ui.adsabs.harvard.edu/abs/2022AJ....163..235L}
}

@ARTICLE{2015NewA...34..262H,
       author = {{Hasanzadeh}, A. and {Farsian}, F. and {Nemati}, M.},
        title = "{New light curve analysis and period changes of the overcontact binary EQ Tauri}",
      journal = {\na},
         year = 2015,
        month = jan,
       volume = {34},
        pages = {262-265},
          doi = {10.1016/j.newast.2014.07.013},
       adsurl = {https://ui.adsabs.harvard.edu/abs/2015NewA...34..262H}
}

@ARTICLE{2020PASJ...72...73L,
       author = {{Liu}, Nian-Ping and {Sarotsakulchai}, Thawicharat and {Rattanasoon}, Somsawat and {Zhang}, Bin},
        title = "{Comprehensive photometric investigation of an active early K-type contact system{\textemdash}IL Cancri}",
      journal = {\pasj},
         year = 2020,
        month = oct,
       volume = {72},
       number = {5},
          eid = {73},
        pages = {73},
          doi = {10.1093/pasj/psaa062},
archivePrefix = {arXiv},
       eprint = {2006.12711},
 primaryClass = {astro-ph.SR},
       adsurl = {https://ui.adsabs.harvard.edu/abs/2020PASJ...72...73L}
}

@ARTICLE{2025AJ....169..139X,
       author = {{Xia}, Qiqi and {Wang}, Xiaofeng and {Li}, Kai and {Gao}, Xiang and {Guo}, Fangzhou and {Lin}, Jie and {Liu}, Cheng and {Mo}, Jun and {Peng}, Haowei and {Liu}, Qichun and {Xi}, Gaobo and {Yan}, Shengyu and {Jiang}, Xiaojun and {Zhang}, Jicheng and {Song}, Cui-Ying and {Shi}, Jianrong and {Ma}, Xiaoran and {Xiang}, Danfeng and {Li}, Wenxiong},
        title = "{Minute-cadence Observations of the LAMOST Fields with the TMTS. VI. Physical Parameters of Contact Binaries}",
      journal = {\aj},
         year = 2025,
        month = mar,
       volume = {169},
       number = {3},
          eid = {139},
        pages = {139},
          doi = {10.3847/1538-3881/ada7eb},
archivePrefix = {arXiv},
       eprint = {2412.11545},
 primaryClass = {astro-ph.SR},
       adsurl = {https://ui.adsabs.harvard.edu/abs/2025AJ....169..139X}
}

@ARTICLE{2018MNRAS.477.3145J,
       author = {{Jayasinghe}, T. and {Kochanek}, C.~S. and {Stanek}, K.~Z. and {Shappee}, B.~J. and {Holoien}, T.~W. -S. and {Thompson}, Todd A. and {Prieto}, J.~L. and {Dong}, Subo and {Pawlak}, M. and {Shields}, J.~V. and {Pojmanski}, G. and {Otero}, S. and {Britt}, C.~A. and {Will}, D.},
        title = "{The ASAS-SN catalogue of variable stars I: The Serendipitous Survey}",
      journal = {\mnras},
         year = 2018,
        month = jul,
       volume = {477},
       number = {3},
        pages = {3145-3163},
          doi = {10.1093/mnras/sty838},
archivePrefix = {arXiv},
       eprint = {1803.01001},
 primaryClass = {astro-ph.SR},
       adsurl = {https://ui.adsabs.harvard.edu/abs/2018MNRAS.477.3145J}
}

@ARTICLE{2021AJ....162...13L,
       author = {{Li}, Kai and {Xia}, Qi-Qi and {Kim}, Chun-Hwey and {Gao}, Xing and {Hu}, Shao-Ming and {Guo}, Di-Fu and {Gao}, Dong-Yang and {Chen}, Xu and {Guo}, Ya-Ni},
        title = "{Photometric Study and Absolute Parameter Estimation of Six Totally Eclipsing Contact Binaries}",
      journal = {\aj},
         year = 2021,
        month = jul,
       volume = {162},
       number = {1},
          eid = {13},
        pages = {13},
          doi = {10.3847/1538-3881/abfc53},
archivePrefix = {arXiv},
       eprint = {2104.13759},
 primaryClass = {astro-ph.SR},
       adsurl = {https://ui.adsabs.harvard.edu/abs/2021AJ....162...13L}
}

@ARTICLE{2012RAA....12.1197C,
       author = {{Cui}, Xiang-Qun and {Zhao}, Yong-Heng and {Chu}, Yao-Quan and {Li}, Guo-Ping and {Li}, Qi and {Zhang}, Li-Ping and {Su}, Hong-Jun and {Yao}, Zheng-Qiu and {Wang}, Ya-Nan and {Xing}, Xiao-Zheng and {Li}, Xin-Nan and {Zhu}, Yong-Tian and {Wang}, Gang and {Gu}, Bo-Zhong and {Luo}, A. -Li and {Xu}, Xin-Qi and {Zhang}, Zhen-Chao and {Liu}, Gen-Rong and {Zhang}, Hao-Tong and {Yang}, De-Hua and {Cao}, Shu-Yun and {Chen}, Hai-Yuan and {Chen}, Jian-Jun and {Chen}, Kun-Xin and {Chen}, Ying and {Chu}, Jia-Ru and {Feng}, Lei and {Gong}, Xue-Fei and {Hou}, Yong-Hui and {Hu}, Hong-Zhuan and {Hu}, Ning-Sheng and {Hu}, Zhong-Wen and {Jia}, Lei and {Jiang}, Fang-Hua and {Jiang}, Xiang and {Jiang}, Zi-Bo and {Jin}, Ge and {Li}, Ai-Hua and {Li}, Yan and {Li}, Ye-Ping and {Liu}, Guan-Qun and {Liu}, Zhi-Gang and {Lu}, Wen-Zhi and {Mao}, Yin-Dun and {Men}, Li and {Qi}, Yong-Jun and {Qi}, Zhao-Xiang and {Shi}, Huo-Ming and {Tang}, Zheng-Hong and {Tao}, Qing-Sheng and {Wang}, Da-Qi and {Wang}, Dan and {Wang}, Guo-Min and {Wang}, Hai and {Wang}, Jia-Ning and {Wang}, Jian and {Wang}, Jian-Ling and {Wang}, Jian-Ping and {Wang}, Lei and {Wang}, Shu-Qing and {Wang}, You and {Wang}, Yue-Fei and {Xu}, Ling-Zhe and {Xu}, Yan and {Yang}, Shi-Hai and {Yu}, Yong and {Yuan}, Hui and {Yuan}, Xiang-Yan and {Zhai}, Chao and {Zhang}, Jing and {Zhang}, Yan-Xia and {Zhang}, Yong and {Zhao}, Ming and {Zhou}, Fang and {Zhou}, Guo-Hua and {Zhu}, Jie and {Zou}, Si-Cheng},
        title = "{The Large Sky Area Multi-Object Fiber Spectroscopic Telescope (LAMOST)}",
      journal = {Research in Astronomy and Astrophysics},
         year = 2012,
        month = sep,
       volume = {12},
       number = {9},
        pages = {1197-1242},
          doi = {10.1088/1674-4527/12/9/003},
       adsurl = {https://ui.adsabs.harvard.edu/abs/2012RAA....12.1197C}
}

@ARTICLE{2015RAA....15.1095L,
       author = {{Luo}, A. -Li and {Zhao}, Yong-Heng and {Zhao}, Gang and {Deng}, Li-Cai and {Liu}, Xiao-Wei and {Jing}, Yi-Peng and {Wang}, Gang and {Zhang}, Hao-Tong and {Shi}, Jian-Rong and {Cui}, Xiang-Qun and {Chu}, Yao-Quan and {Li}, Guo-Ping and {Bai}, Zhong-Rui and {Wu}, Yue and {Cai}, Yan and {Cao}, Shu-Yun and {Cao}, Zi-Huang and {Carlin}, Jeffrey L. and {Chen}, Hai-Yuan and {Chen}, Jian-Jun and {Chen}, Kun-Xin and {Chen}, Li and {Chen}, Xue-Lei and {Chen}, Xiao-Yan and {Chen}, Ying and {Christlieb}, Norbert and {Chu}, Jia-Ru and {Cui}, Chen-Zhou and {Dong}, Yi-Qiao and {Du}, Bing and {Fan}, Dong-Wei and {Feng}, Lei and {Fu}, Jian-Ning and {Gao}, Peng and {Gong}, Xue-Fei and {Gu}, Bo-Zhong and {Guo}, Yan-Xin and {Han}, Zhan-Wen and {He}, Bo-Liang and {Hou}, Jin-Liang and {Hou}, Yong-Hui and {Hou}, Wen and {Hu}, Hong-Zhuan and {Hu}, Ning-Sheng and {Hu}, Zhong-Wen and {Huo}, Zhi-Ying and {Jia}, Lei and {Jiang}, Fang-Hua and {Jiang}, Xiang and {Jiang}, Zhi-Bo and {Jin}, Ge and {Kong}, Xiao and {Kong}, Xu and {Lei}, Ya-Juan and {Li}, Ai-Hua and {Li}, Chang-Hua and {Li}, Guang-Wei and {Li}, Hai-Ning and {Li}, Jian and {Li}, Qi and {Li}, Shuang and {Li}, Sha-Sha and {Li}, Xin-Nan and {Li}, Yan and {Li}, Yin-Bi and {Li}, Ye-Ping and {Liang}, Yuan and {Lin}, Chien-Cheng and {Liu}, Chao and {Liu}, Gen-Rong and {Liu}, Guan-Qun and {Liu}, Zhi-Gang and {Lu}, Wen-Zhi and {Luo}, Yu and {Mao}, Yin-Dun and {Newberg}, Heidi and {Ni}, Ji-Jun and {Qi}, Zhao-Xiang and {Qi}, Yong-Jun and {Shen}, Shi-Yin and {Shi}, Huo-Ming and {Song}, Jing and {Song}, Yi-Han and {Su}, Ding-Qiang and {Su}, Hong-Jun and {Tang}, Zheng-Hong and {Tao}, Qing-Sheng and {Tian}, Yuan and {Wang}, Dan and {Wang}, Da-Qi and {Wang}, Feng-Fei and {Wang}, Guo-Min and {Wang}, Hai and {Wang}, Hong-Chi and {Wang}, Jian and {Wang}, Jia-Ning and {Wang}, Jian-Ling and {Wang}, Jian-Ping and {Wang}, Jun-Xian and {Wang}, Lei and {Wang}, Meng-Xin and {Wang}, Shou-Guan and {Wang}, Shu-Qing and {Wang}, Xia and {Wang}, Ya-Nan and {Wang}, You and {Wang}, Yue-Fei and {Wang}, You-Fen and {Wei}, Peng and {Wei}, Ming-Zhi and {Wu}, Hong and {Wu}, Ke-Fei and {Wu}, Xue-Bing and {Wu}, Yu-Zhong and {Xing}, Xiao-Zheng and {Xu}, Ling-Zhe and {Xu}, Xin-Qi and {Xu}, Yan and {Yan}, Tai-Sheng and {Yang}, De-Hua and {Yang}, Hai-Feng and {Yang}, Hui-Qin and {Yang}, Ming and {Yao}, Zheng-Qiu and {Yu}, Yong and {Yuan}, Hui and {Yuan}, Hai-Bo and {Yuan}, Hai-Long and {Yuan}, Wei-Min and {Zhai}, Chao and {Zhang}, En-Peng and {Zhang}, Hua-Wei and {Zhang}, Jian-Nan and {Zhang}, Li-Pin and {Zhang}, Wei and {Zhang}, Yong and {Zhang}, Yan-Xia and {Zhang}, Zheng-Chao and {Zhao}, Ming and {Zhou}, Fang and {Zhou}, Xu and {Zhu}, Jie and {Zhu}, Yong-Tian and {Zou}, Si-Cheng and {Zuo}, Fang},
        title = "{The first data release (DR1) of the LAMOST regular survey}",
      journal = {Research in Astronomy and Astrophysics},
         year = 2015,
        month = aug,
       volume = {15},
       number = {8},
          eid = {1095},
        pages = {1095},
          doi = {10.1088/1674-4527/15/8/002},
archivePrefix = {arXiv},
       eprint = {1505.01570},
 primaryClass = {astro-ph.GA},
       adsurl = {https://ui.adsabs.harvard.edu/abs/2015RAA....15.1095L}
}

@ARTICLE{1971ApJ...166..605W,
       author = {{Wilson}, Robert E. and {Devinney}, Edward J.},
        title = "{Realization of Accurate Close-Binary Light Curves: Application to MR Cygni}",
      journal = {\apj},
         year = 1971,
        month = jun,
       volume = {166},
        pages = {605},
          doi = {10.1086/150986},
       adsurl = {https://ui.adsabs.harvard.edu/abs/1971ApJ...166..605W}
}

@ARTICLE{1979ApJ...234.1054W,
       author = {{Wilson}, R.~E.},
        title = "{Eccentric orbit generalization and simultaneous solution of binary star light and velocity curves.}",
      journal = {\apj},
         year = 1979,
        month = dec,
       volume = {234},
        pages = {1054-1066},
          doi = {10.1086/157588},
       adsurl = {https://ui.adsabs.harvard.edu/abs/1979ApJ...234.1054W}
}

@ARTICLE{1990ApJ...356..613W,
       author = {{Wilson}, R.~E.},
        title = "{Accuracy and Efficiency in the Binary Star Reflection Effect}",
      journal = {\apj},
         year = 1990,
        month = jun,
       volume = {356},
        pages = {613},
          doi = {10.1086/168867},
       adsurl = {https://ui.adsabs.harvard.edu/abs/1990ApJ...356..613W}
}

@ARTICLE{1924MNRAS..84..665V,
       author = {{von Zeipel}, H.},
        title = "{The radiative equilibrium of a rotating system of gaseous masses}",
      journal = {\mnras},
         year = 1924,
        month = jun,
       volume = {84},
        pages = {665-683},
          doi = {10.1093/mnras/84.9.665},
       adsurl = {https://ui.adsabs.harvard.edu/abs/1924MNRAS..84..665V}
}

@ARTICLE{1993AJ....106.2096V,
       author = {{van Hamme}, W.},
        title = "{New Limb-Darkening Coefficients for Modeling Binary Star Light Curves}",
      journal = {\aj},
         year = 1993,
        month = nov,
       volume = {106},
        pages = {2096},
          doi = {10.1086/116788},
       adsurl = {https://ui.adsabs.harvard.edu/abs/1993AJ....106.2096V}
}

@ARTICLE{2019MNRAS.485.4588L,
       author = {{Li}, Kai and {Xia}, Qi-Qi and {Michel}, Raul and {Hu}, Shao-Ming and {Guo}, Di-Fu and {Gao}, Xing and {Chen}, Xu and {Gao}, Dong-Yang},
        title = "{Contact binaries at the short period cut-off - I. Statistics and the first photometric investigations of 10 totally eclipsing systems}",
      journal = {\mnras},
         year = 2019,
        month = jun,
       volume = {485},
       number = {4},
        pages = {4588-4600},
          doi = {10.1093/mnras/stz715},
archivePrefix = {arXiv},
       eprint = {1903.04765},
 primaryClass = {astro-ph.SR},
       adsurl = {https://ui.adsabs.harvard.edu/abs/2019MNRAS.485.4588L}
}

@ARTICLE{1951PRCO....2...85O,
       author = {{O'Connell}, D.~J.~K.},
        title = "{The so-called periastron effect in close eclipsing binaries ; New variable stars (fifth list)}",
      journal = {Publications of the Riverview College Observatory},
         year = 1951,
        month = aug,
       volume = {2},
       number = {6},
        pages = {85-100},
       adsurl = {https://ui.adsabs.harvard.edu/abs/1951PRCO....2...85O}
}

@ARTICLE{2013AJ....146..157C,
       author = {{Christopoulou}, P. -E. and {Papageorgiou}, A.},
        title = "{An Extensive Analysis of the Triple W UMa Type Binary FI Boo}",
      journal = {\aj},
         year = 2013,
        month = dec,
       volume = {146},
       number = {6},
          eid = {157},
        pages = {157},
          doi = {10.1088/0004-6256/146/6/157},
       adsurl = {https://ui.adsabs.harvard.edu/abs/2013AJ....146..157C}
}

@ARTICLE{1967ZA.....65...89L,
       author = {{Lucy}, L.~B.},
        title = "{Gravity-Darkening for Stars with Convective Envelopes}",
      journal = {\zap},
         year = 1967,
        month = jan,
       volume = {65},
        pages = {89},
       adsurl = {https://ui.adsabs.harvard.edu/abs/1967ZA.....65...89L}
}

@ARTICLE{2025ApJ...979...69W,
       author = {{Wang}, Jing-Yi and {Li}, Kai and {Gao}, Xiang and {Guo}, Di-Fu and {Wang}, Li-Heng and {Gao}, Dong-Yang and {Li}, Ling-Zhi and {Guo}, Ya-Ni and {Gao}, Xing and {Sun}, Guo-You},
        title = "{Search for and Analysis of Eclipsing Binaries in the LAMOST Medium-resolution Survey Field. I. R.A.: 23$^{h}$01$^{m}$51$^{s}$, Decl.: +34{\textdegree}36'45″}",
      journal = {\apj},
         year = 2025,
        month = jan,
       volume = {979},
       number = {1},
          eid = {69},
        pages = {69},
          doi = {10.3847/1538-4357/ad9929},
archivePrefix = {arXiv},
       eprint = {2412.00377},
 primaryClass = {astro-ph.SR},
       adsurl = {https://ui.adsabs.harvard.edu/abs/2025ApJ...979...69W}
}

@ARTICLE{1969AcA....19..245R,
       author = {{Ruci{\'n}ski}, S.~M.},
        title = "{The Proximity Effects in Close Binary Systems. II. The Bolometric Reflection Effect for Stars with Deep Convective Envelopes}",
      journal = {\actaa},
         year = 1969,
        month = jan,
       volume = {19},
        pages = {245},
       adsurl = {https://ui.adsabs.harvard.edu/abs/1969AcA....19..245R}
}

@ARTICLE{2020RNAAS...4..201C,
       author = {{Caldwell}, Douglas A. and {Tenenbaum}, Peter and {Twicken}, Joseph D. and {Jenkins}, Jon M. and {Ting}, Eric and {Smith}, Jeffrey C. and {Hedges}, Christina and {Fausnaugh}, Michael M. and {Rose}, Mark and {Burke}, Christopher},
        title = "{TESS Science Processing Operations Center FFI Target List Products}",
      journal = {Research Notes of the American Astronomical Society},
         year = 2020,
        month = nov,
       volume = {4},
       number = {11},
          eid = {201},
        pages = {201},
          doi = {10.3847/2515-5172/abc9b3},
archivePrefix = {arXiv},
       eprint = {2011.05495},
 primaryClass = {astro-ph.EP},
       adsurl = {https://ui.adsabs.harvard.edu/abs/2020RNAAS...4..201C}
}

@ARTICLE{2015JATIS...1a4003R,
       author = {{Ricker}, George R. and {Winn}, Joshua N. and {Vanderspek}, Roland and {Latham}, David W. and {Bakos}, G{\'a}sp{\'a}r {\'A}. and {Bean}, Jacob L. and {Berta-Thompson}, Zachory K. and {Brown}, Timothy M. and {Buchhave}, Lars and {Butler}, Nathaniel R. and {Butler}, R. Paul and {Chaplin}, William J. and {Charbonneau}, David and {Christensen-Dalsgaard}, J{\o}rgen and {Clampin}, Mark and {Deming}, Drake and {Doty}, John and {De Lee}, Nathan and {Dressing}, Courtney and {Dunham}, Edward W. and {Endl}, Michael and {Fressin}, Francois and {Ge}, Jian and {Henning}, Thomas and {Holman}, Matthew J. and {Howard}, Andrew W. and {Ida}, Shigeru and {Jenkins}, Jon M. and {Jernigan}, Garrett and {Johnson}, John Asher and {Kaltenegger}, Lisa and {Kawai}, Nobuyuki and {Kjeldsen}, Hans and {Laughlin}, Gregory and {Levine}, Alan M. and {Lin}, Douglas and {Lissauer}, Jack J. and {MacQueen}, Phillip and {Marcy}, Geoffrey and {McCullough}, Peter R. and {Morton}, Timothy D. and {Narita}, Norio and {Paegert}, Martin and {Palle}, Enric and {Pepe}, Francesco and {Pepper}, Joshua and {Quirrenbach}, Andreas and {Rinehart}, Stephen A. and {Sasselov}, Dimitar and {Sato}, Bun'ei and {Seager}, Sara and {Sozzetti}, Alessandro and {Stassun}, Keivan G. and {Sullivan}, Peter and {Szentgyorgyi}, Andrew and {Torres}, Guillermo and {Udry}, Stephane and {Villasenor}, Joel},
        title = "{Transiting Exoplanet Survey Satellite (TESS)}",
      journal = {Journal of Astronomical Telescopes, Instruments, and Systems},
         year = 2015,
        month = jan,
       volume = {1},
          eid = {014003},
        pages = {014003},
          doi = {10.1117/1.JATIS.1.1.014003},
       adsurl = {https://ui.adsabs.harvard.edu/abs/2015JATIS...1a4003R}
}

@ARTICLE{1998AJ....116.2998R,
       author = {{Rucinski}, Slavek M.},
        title = "{Contact Binaries of the Galactic Disk: Comparison of the Baade's Window and Open Cluster Samples}",
      journal = {\aj},
         year = 1998,
        month = dec,
       volume = {116},
       number = {6},
        pages = {2998-3017},
          doi = {10.1086/300644},
archivePrefix = {arXiv},
       eprint = {astro-ph/9806154},
 primaryClass = {astro-ph},
       adsurl = {https://ui.adsabs.harvard.edu/abs/1998AJ....116.2998R}
}

@ARTICLE{2017AJ....154..125M,
       author = {{Mateo}, Nicole M. and {Rucinski}, Slavek M.},
        title = "{Absolute-magnitude Calibration for W UMa-type Systems Based on Gaia Data}",
      journal = {\aj},
         year = 2017,
        month = sep,
       volume = {154},
       number = {3},
          eid = {125},
        pages = {125},
          doi = {10.3847/1538-3881/aa8453},
archivePrefix = {arXiv},
       eprint = {1708.01097},
 primaryClass = {astro-ph.SR},
       adsurl = {https://ui.adsabs.harvard.edu/abs/2017AJ....154..125M}
}

@ARTICLE{2002MNRAS.329..897H,
       author = {{Hurley}, Jarrod R. and {Tout}, Christopher A. and {Pols}, Onno R.},
        title = "{Evolution of binary stars and the effect of tides on binary populations}",
      journal = {\mnras},
         year = 2002,
        month = feb,
       volume = {329},
       number = {4},
        pages = {897-928},
          doi = {10.1046/j.1365-8711.2002.05038.x},
archivePrefix = {arXiv},
       eprint = {astro-ph/0201220},
 primaryClass = {astro-ph},
       adsurl = {https://ui.adsabs.harvard.edu/abs/2002MNRAS.329..897H}
}

@ARTICLE{2013MNRAS.430.2029Y,
       author = {{Yildiz}, M. and {Do{\u{g}}an}, T.},
        title = "{On the origin of W UMa type contact binaries - a new method for computation of initial masses}",
      journal = {\mnras},
         year = 2013,
        month = apr,
       volume = {430},
       number = {3},
        pages = {2029-2038},
          doi = {10.1093/mnras/stt028},
archivePrefix = {arXiv},
       eprint = {1301.6035},
 primaryClass = {astro-ph.SR},
       adsurl = {https://ui.adsabs.harvard.edu/abs/2013MNRAS.430.2029Y}
}

@ARTICLE{2004A&A...426.1001C,
	author = {{Csizmadia}, Sz. and {Klagyivik}, P.},
	title = "{On the properties of contact binary stars}",
	journal = {\aap},
	year = 2004,
	month = nov,
	volume = {426},
	pages = {1001-1005},
	doi = {10.1051/0004-6361:20040430},
	archivePrefix = {arXiv},
	eprint = {astro-ph/0408049},
	primaryClass = {astro-ph},
	adsurl = {https://ui.adsabs.harvard.edu/abs/2004A&A...426.1001C}
}

@ARTICLE{2025AJ....169...85X,
       author = {{Xu}, Xin and {Li}, Kai and {Liu}, Fei and {Yan}, Qian-Xue and {Wang}, Yi-Fan and {Cui}, Xin-Yu and {Wang}, Jing-Yi and {Gao}, Xing and {Sun}, Guo-You and {Wu}, Cheng-Yu and {Li}, Mu-Zi-Mei},
        title = "{Photometric and Spectroscopic Investigations of Three Large Amplitude Contact Binaries}",
      journal = {\aj},
         year = 2025,
        month = feb,
       volume = {169},
       number = {2},
          eid = {85},
        pages = {85},
          doi = {10.3847/1538-3881/ad9a59},
archivePrefix = {arXiv},
       eprint = {2412.00331},
 primaryClass = {astro-ph.SR},
       adsurl = {https://ui.adsabs.harvard.edu/abs/2025AJ....169...85X}
}

@ARTICLE{2023AJ....166..200C,
       author = {{Cook}, Evan M. and {Kobulnicky}, Henry A.},
        title = "{Observational Constraints on Close Binary Star Evolution. I. Putative Contact Binaries with Long Periods and High Mass Ratios}",
      journal = {\aj},
         year = 2023,
        month = nov,
       volume = {166},
       number = {5},
          eid = {200},
        pages = {200},
          doi = {10.3847/1538-3881/acfc47},
       adsurl = {https://ui.adsabs.harvard.edu/abs/2023AJ....166..200C}
}

@ARTICLE{2025AJ....170..167L,
       author = {{Larsen}, Conor M. and {Pr{\v{s}}a}, Andrej},
        title = "{Binary Analysis and Period Study of the Long-period, High-mass-ratio Contact Binary KIC 7766185}",
      journal = {\aj},
         year = 2025,
        month = sep,
       volume = {170},
       number = {3},
          eid = {167},
        pages = {167},
          doi = {10.3847/1538-3881/adf202},
archivePrefix = {arXiv},
       eprint = {2507.14979},
 primaryClass = {astro-ph.SR},
       adsurl = {https://ui.adsabs.harvard.edu/abs/2025AJ....170..167L}
}

@ARTICLE{2025AJ....170..214P,
       author = {{Poro}, Atila and {Li}, Kai and {Michel}, Raul and {Wang}, Li-Heng and {Alicavus}, Fahri and {N{\'a}jera}, Morgan Rhai and {Santill{\'a}n-Ortega}, Priscila and {Tamayo}, Francisco Javier and {Aceves}, Hector},
        title = "{BSN. II. The First Light Curve Study of Eight Total Eclipsing Contact Binary Stars with Shallow Fillout Factors}",
      journal = {\aj},
         year = 2025,
        month = oct,
       volume = {170},
       number = {4},
          eid = {214},
        pages = {214},
          doi = {10.3847/1538-3881/adfc57},
archivePrefix = {arXiv},
       eprint = {2508.11901},
 primaryClass = {astro-ph.SR},
       adsurl = {https://ui.adsabs.harvard.edu/abs/2025AJ....170..214P}
}

@ARTICLE{2015AJ....150...69Y,
       author = {{Yang}, Yuan-Gui and {Qian}, Sheng-Bang},
        title = "{Deep, Low Mass Ratio Overcontact Binary Systems. XIV. A Statistical Analysis of 46 Sample Binaries}",
      journal = {\aj},
         year = 2015,
        month = sep,
       volume = {150},
       number = {3},
          eid = {69},
        pages = {69},
          doi = {10.1088/0004-6256/150/3/69},
       adsurl = {https://ui.adsabs.harvard.edu/abs/2015AJ....150...69Y}
}

@ARTICLE{2025MNRAS.541.3401Z,
       author = {{Zhou}, Xiao},
        title = "{An extremely low mass ratio contact binary NW Aps with a potential compact companion star}",
      journal = {\mnras},
         year = 2025,
        month = aug,
       volume = {541},
       number = {4},
        pages = {3401-3411},
          doi = {10.1093/mnras/staf1170},
archivePrefix = {arXiv},
       eprint = {2507.17198},
 primaryClass = {astro-ph.SR},
       adsurl = {https://ui.adsabs.harvard.edu/abs/2025MNRAS.541.3401Z}
}

@ARTICLE{2006MNRAS.373.1483E,
       author = {{Eker}, Z. and {Demircan}, O. and {Bilir}, S. and {Karata{\c{s}}}, Y.},
        title = "{Dynamical evolution of active detached binaries on the logJ$_{o}$-logM diagram and contact binary formation}",
      journal = {\mnras},
         year = 2006,
        month = dec,
       volume = {373},
       number = {4},
        pages = {1483-1494},
          doi = {10.1111/j.1365-2966.2006.11073.x},
archivePrefix = {arXiv},
       eprint = {astro-ph/0609395},
 primaryClass = {astro-ph},
       adsurl = {https://ui.adsabs.harvard.edu/abs/2006MNRAS.373.1483E}
}

@ARTICLE{1968ApJ...151.1123L,
       author = {{Lucy}, L.~B.},
        title = "{The Structure of Contact Binaries}",
      journal = {\apj},
         year = 1968,
        month = mar,
       volume = {151},
        pages = {1123},
          doi = {10.1086/149510},
       adsurl = {https://ui.adsabs.harvard.edu/abs/1968ApJ...151.1123L}
}

@ARTICLE{2016RAA....16...68Z,
       author = {{Zhu}, Li-Ying and {Zhao}, Er-Gang and {Zhou}, Xiao},
        title = "{A low-mass-ratio and deep contact binary as the progenitor of the merger V1309 Sco}",
      journal = {Research in Astronomy and Astrophysics},
         year = 2016,
        month = apr,
       volume = {16},
       number = {4},
          eid = {68},
        pages = {68},
          doi = {10.1088/1674-4527/16/4/068},
archivePrefix = {arXiv},
       eprint = {1611.04699},
 primaryClass = {astro-ph.SR},
       adsurl = {https://ui.adsabs.harvard.edu/abs/2016RAA....16...68Z}
}

@ARTICLE{2003ChJAA...3..142L,
       author = {{Liu}, Qing-Yao and {Yang}, Yu-Lan},
        title = "{A Possible Explanation of the O'Connell Effect in Close Binary Stars}",
      journal = {\cjaa},
         year = 2003,
        month = apr,
       volume = {3},
        pages = {142-150},
          doi = {10.1088/1009-9271/3/2/142},
       adsurl = {https://ui.adsabs.harvard.edu/abs/2003ChJAA...3..142L}
}

@ARTICLE{1967MmRAS..70..111E,
       author = {{Eggen}, O.~J.},
        title = "{Contact binaries, II.}",
      journal = {\memras},
         year = 1967,
        month = jan,
       volume = {70},
        pages = {111},
       adsurl = {https://ui.adsabs.harvard.edu/abs/1967MmRAS..70..111E}
}

@ARTICLE{1994PASP..106..462R,
       author = {{Rucinski}, S.~M.},
        title = "{M\_V = M\_V(Log P, Log T\_e) Calibrations for W UMa Systems}",
      journal = {\pasp},
         year = 1994,
        month = may,
       volume = {106},
        pages = {462},
          doi = {10.1086/133401},
       adsurl = {https://ui.adsabs.harvard.edu/abs/1994PASP..106..462R}
}

@ARTICLE{1976ApJ...205..217F,
       author = {{Flannery}, Brian P.},
        title = "{A Cyclic Thermal Instability in Contact Binary Stars}",
      journal = {\apj},
         year = 1976,
        month = apr,
       volume = {205},
        pages = {217-225},
          doi = {10.1086/154266},
       adsurl = {https://ui.adsabs.harvard.edu/abs/1976ApJ...205..217F}
}
\bibliographystyle{aasjournal}



\end{document}